\documentclass{IEEEtran}
\usepackage{cite}
\usepackage{amsmath,amssymb,amsfonts}
\usepackage{enumerate}
\usepackage{stmaryrd}
\usepackage{multirow}
\usepackage{algorithmic}
\usepackage{graphicx}
\usepackage[hidelinks]{hyperref}
\usepackage{graphics}
\usepackage{textcomp}
\usepackage{algorithm}
\usepackage{booktabs}
\usepackage{tikz}
\usepackage{color}
\usepackage{bm}
\usepackage{epstopdf}
\usepackage{cases}
\usepackage{subfigure}

\def\BibTeX{{\rm B\kern-.05em{\sc i\kern-.025em b}\kern-.08em T\kern-.1667em\lower.7ex\hbox{E}\kern-.125emX}}
\begin{document}
\title{Non-Uniform Antenna Array Design with Large Inter-Element Spacing for Massive MIMO}
\author{Lin Chen, Qu Luo,~\IEEEmembership{Member,~IEEE}, Orestis Christogeorgos, Pei Xiao,~\IEEEmembership{Senior Member,~IEEE}, Gabriele Gradoni,~\IEEEmembership{Senior Member,~IEEE}, Mohsen Khalily,~\IEEEmembership{Senior Member,~IEEE}, Yang Hao,~\IEEEmembership{Fellow,~IEEE}, and Hongbin Li,~\IEEEmembership{Fellow,~IEEE}
\thanks{Lin Chen and Hongbin Li are with the Department of Electrical and Computer Engineering, Stevens Institute of Technology, Hoboken, NJ 07030, USA (e-mail: lchen53@stevens.edu; hli@stevens.edu). Qu Luo, Pei Xiao, Gabriele Gradoni, and Mohsen Khalily are with the Institute for Communication Systems (ICS), Home of the 5GIC \& 6GIC, University of Surrey, Guildford GU2 7XH, U.K. (e-mail: q.u.luo@surrey.ac.uk; p.xiao@surrey.ac.uk; g.gradoni@surrey.ac.uk; m.khalily@surrey.ac.uk). Orestis Christogeorgos and Yang Hao are with the School of Electronic Engineering and Computer Science, Queen Mary University of London, E1 4NS London, U.K. (e-mail: o.christogeorgos@qmul.ac.uk; y.hao@qmul.ac.uk). (Corresponding author: Hongbin Li)}}
\maketitle
\begin{abstract}
In massive multiple-input multiple-output (MIMO) systems, uniform arrays are typically configured with inter-element spacing no greater than half a wavelength to avoid grating lobes and spatial aliasing. However, many emerging fifth- and sixth-generation (5G/6G) applications rely on distributed arrays whose inter-element spacing far exceeds half a wavelength. In this paper, we propose an electromagnetic mutual-information-theoretic (EMIT)-guided non-uniform array (NUA) design with large inter-element spacing for massive MIMO systems to address the grating lobes and spatial aliasing artifacts, and in the meantime, to reduce the hardware cost and energy consumption. We start by developing a multipath channel model for non-uniform planar arrays, and analyze the resulting channel characteristics in terms of inter-user interference, aperture efficiency, favorable propagation and channel capacity for the proposed typical NUA patterns. The model is further extended to wideband scenarios, where NUAs demonstrate improved robustness against beam squint due to their more compact element distribution. In addition, we introduce an EMIT approach to NUA design,  which links the spatial sampling pattern of an antenna array to the capacity of the resulting MIMO channel. This gives rise to two complementary shaping strategies, amplitude tapering and geometric shaping, and their joint optimization. Numerical results demonstrate that the proposed NUAs significantly outperform conventional uniform arrays in aperture efficiency, channel orthogonality, beam squint mitigation, capacity, and error rate performance.
\end{abstract}
\section{Introduction}
\IEEEPARstart{T}{he} Nyquist-Shannon sampling theorem \cite{Shannon1,Shannon3} is a fundamental principle for signal processing and communications, providing a sufficient condition for sampling and reconstructing signals without loss of information. Specifically, it dictates that the sampling rate must be at least twice that of the signal bandwidth to avoid aliasing distortion. In the spatial domain, a continuous signal can be perfectly reconstructed from its samples if the spatial sampling interval (i.e., inter-element spacing in a uniformly distributed array) is smaller than half the wavelength of the highest frequency component in the signal.
\par A prominent application of the spatial Nyquist sampling criterion is in massive multiple-input multiple-output (MIMO) systems \cite{MIMOAL}.
Conventional  MIMO  deployments rely on uniform periodic antenna arrays satisfying the spatial Nyquist rate sampling, i.e., inter-element spacing constrained to at most half a wavelength, to avoid spatial aliasing and grating lobes. While this design ensures predictable beam patterns, it imposes three critical bottlenecks for next-generation networks. First, achieving favorable propagation and channel hardening, essential for low-complexity linear detection, requires an ever-increasing number of antennas, escalating hardware complexity and deployment costs \cite{9850359,10552396}. Second, the energy consumption of massive MIMO systems scales with antenna count, conflicting with sustainability goals as wireless traffic continues to grow exponentially. Third, emerging distributed MIMO architectures, e.g., cell-free networks and unmanned aerial vehicle (UAV) swarms, inherently feature large inter-element spacing, and uniform arrays inevitably suffer from grating lobes that cause severe inter-user interference, beam ambiguity, and security vulnerabilities \cite{FP3,11481148}. These limitations raise a fundamental question: \textit{Can we break the half-wavelength spacing constraint while simultaneously reducing antenna count, hardware cost, and energy consumption?}
\par To address these bottlenecks, an active line of research seeks to break the Nyquist spatial sampling limit. Compressive sensing \cite{CS1,CS2,CS3} offers one such approach, which enables the recovery of signals from significantly fewer samples than the Nyquist rate by concentrating signal energy in a few bases and coefficients within a transform domain. However, this approach relies on the premise that the signal of interest is sparse or compressible in a certain domain. In many practical scenarios such as beamforming and wideband array processing, signals may span a wide range of spatial frequencies, making it challenging to find a sparse representation in any particular domain.
\par Inspired by Erwin Schrödinger's concept of the ``aperiodic crystal'' in his famous book ``What Is Life?'' \cite{schrodinger1944life}, we recognize that periodic structures, while orderly, are information-deficient: their repeating patterns cannot encode complex information efficiently. In contrast, aperiodic (non-uniform) arrangements, like DNA's molecular structure, leverage irregular yet structured spatial distributions to maximize information capacity. This biological principle suggests a paradigm shift: rather than treating antenna placement as a uniform geometric constraint, we should view spatial sampling geometry as an information-theoretic degree of freedom for maximizing mutual information between transmitted symbols and received signals. 
\par A recent attempt along this direction is the hyperuniform disordered (HuD) distribution-based array design \cite{HUD,HUD2,HUD3,HUDX,HUD0}, which exploits short-range order and long-range disorder among element positions to suppress grating lobes without requiring signal sparsity or compressibility. The HuD distribution achieves a grating-lobe-free radiation pattern by ensuring that density fluctuations vanish at large scales, thereby breaking the periodicity that gives rise to spatial aliasing. However, the HuD distribution is primarily designed from a statistical mechanics perspective, and its element placement is governed by a stealthiness parameter that controls the degree of structural order \cite{HUD3,HUDX}. As a result, the HuD design does not explicitly account for communication-level performance metrics such as channel capacity or mutual information, leaving room for further optimization from an information-theoretic standpoint.  Nevertheless, the demonstrated success of irregular element distributions in suppressing grating lobes and improving channel orthogonality highlights the significant potential of aperiodic spatial sampling for massive MIMO systems. In this paper, we broadly refer to arrays with aperiodic element distributions as non-uniform arrays (NUAs).
\par
\begin{figure*}[!t]
\subfigure[URA]
{\includegraphics[height=3.6cm]{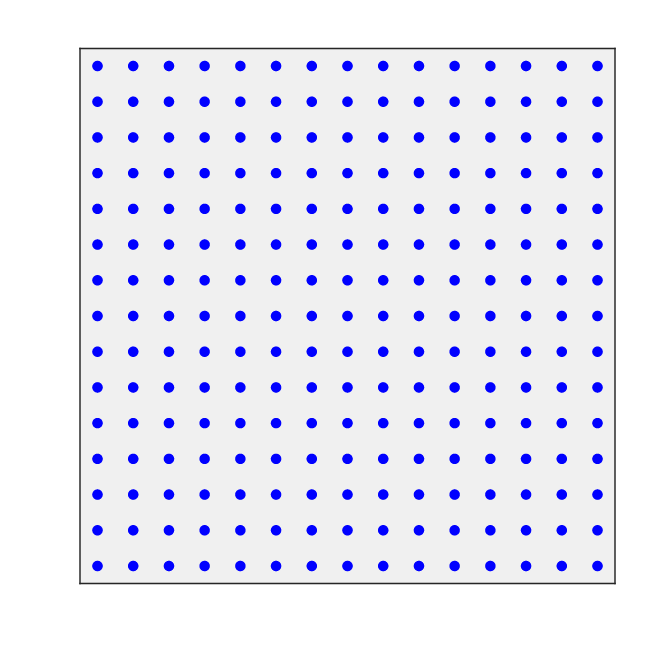}}
\hspace{-0.2cm}
\subfigure[GAM]
{\includegraphics[height=3.6cm]{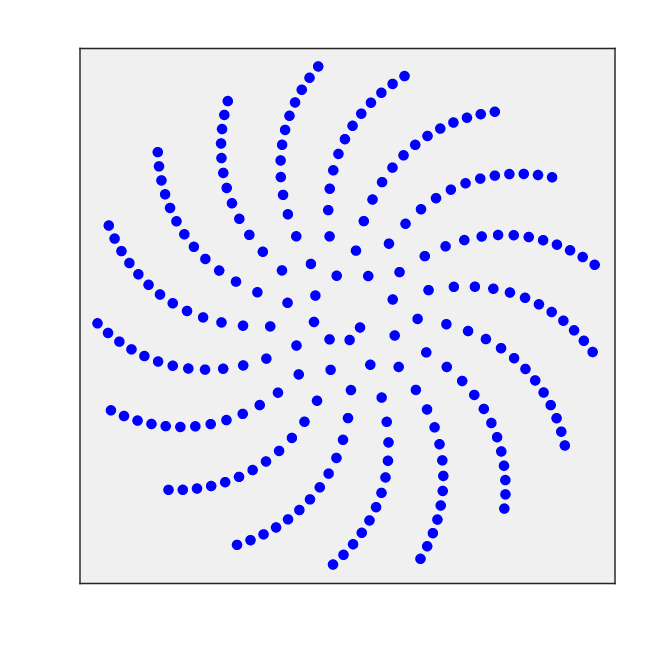}}
\hspace{-0.2cm}
\subfigure[DiscGAM]
{\includegraphics[height=3.6cm]{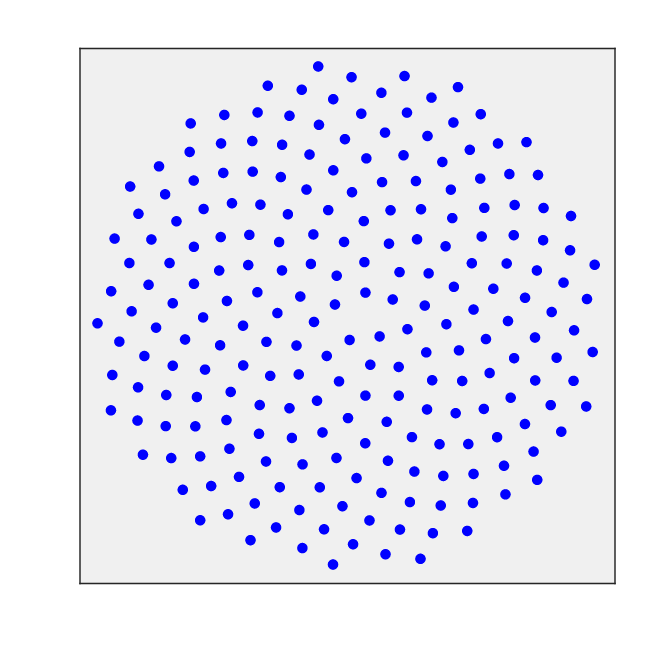}}
\hspace{-0.2cm}
\subfigure[SPM]
{\includegraphics[height=3.6cm]{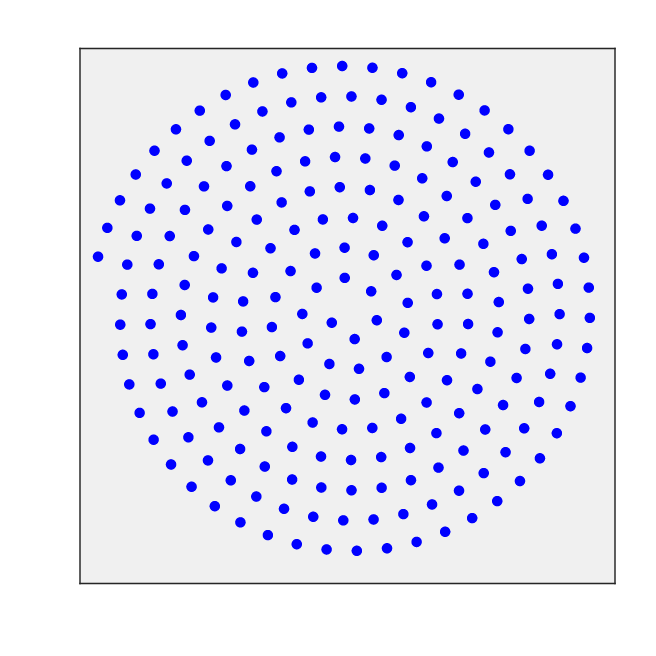}}
\hspace{-0.2cm}
\subfigure[HuD]
{\includegraphics[height=3.6cm]{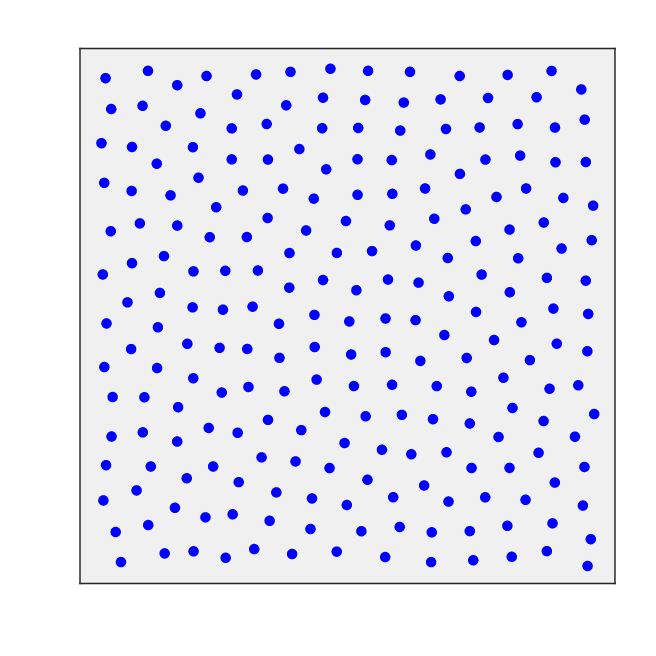}}
\hspace{-0.5cm}
\caption{Illustration of the element pattern for a URA and NUAs with GAM, DiscGAM, SPM, and HuD distributions, each consisting of $M=225$ elements within the same aperture size.}
\label{AllPattern}
\end{figure*}

 Against the above-mentioned background, this paper proposes NUAs to address the limitations of conventional uniform rectangular arrays (URAs) for massive MIMO systems and introduces an electromagnetic mutual-information-theoretic (EMIT) framework for their systematic design. Traditional array design is guided by geometric or field-based criteria such as main-lobe width, sidelobe level, or grating lobe suppression, which only indirectly relate to the fundamental goal of maximizing information throughput. A key insight of this work is that the beam pattern in the angular domain is an invertible Fourier transform of the spatial element distribution. Consequently, any degradation in information transfer must originate from the spatial sampling pattern itself. This allows us to treat the array geometry as a spatial sampling operator and evaluate it rigorously through the capacity of the resulting MIMO channel, rather than relying only on heuristic beam-pattern metrics. The main contributions of this work are summarized as follows.
\begin{itemize}
\item We propose NUAs to overcome the inherent limitations of conventional URAs, particularly the Nyquist spatial sampling constraint imposed by uniform element spacing. To this end, we introduce several NUAs inspired by mutual-information-optimal constellations, including the golden angle modulation (GAM)\cite{GAM,GAM2}, disc-shaped GAM (DiscGAM)\cite{GAM2}, spiral modulation (SPM)\cite{spiral}, and HuD\cite{HUD,HUD2,HUD3,HUDX,HUD0} distributions, as illustrated in Fig. \ref{AllPattern}, to construct flexible spatial sampling patterns. The resulting array configurations are analyzed in terms of channel orthogonality and aperture efficiency, providing insights into their potential performance advantages. Furthermore, the proposed framework is extended to wideband massive-array scenarios, where the results demonstrate that NUAs exhibit enhanced robustness to the beam squint effect \cite{bl,gl,bl2} compared with conventional uniform arrays. This improvement is attributed to their more compact spatial element distributions, which effectively mitigate frequency-dependent beam misalignment.
\item An explicit EMIT approach to NUA design is introduced by treating both the array geometry and the excitation profile as design variables, which naturally leads to the concepts of geometric shaping and amplitude tapering in the spatial domain. To this end, amplitude tapering, geometric shaping, and their joint shaping strategies are proposed to maximize the MIMO channel capacity in the design of NUAs.
\item We conduct extensive simulations to demonstrate the advantages of the proposed NUAs. The results show that typical NUAs exhibit improved channel orthogonality, channel capacity, and bit-error-rate (BER) performance, as well as more favorable propagation conditions, compared with conventional URAs when the inter-element spacing exceeds half a wavelength. Moreover, an NUA with the same array aperture can achieve comparable performance while using a significantly reduced number of antenna elements compared with URAs. These results suggest that NUAs provide a promising solution for next-generation distributed array systems.
\end{itemize}
\par The remainder of this paper is organized as follows. Section \ref{SignalModel} presents the massive MIMO channel model for NUAs in a narrowband scenario. Drawing upon the channel model, Section \ref{advan} investigates the channel characteristics in terms of array factor, channel orthogonality, and favorable propagation. Section \ref{shaping} presents the proposed NUA optimization schemes, i.e., the proposed amplitude tapering and geometric shaping for channel capacity enhancement. Section \ref{beamsquint} extends the proposed NUAs to wideband scenarios. Experimental results in Section \ref{expe} demonstrate the superiority of NUAs over traditional URAs for massive MIMO, highlighting their improved channel orthogonality and system-level performance. Lastly, conclusion and future work are provided in Section \ref{concl}.
\par \emph{Notations}---A scalar is represented by a non-bold lowercase letter $x$ or uppercase letter $X$. A vector and matrix are represented by a bold lowercase letter $\mathbf{x}$ and bold uppercase letter $\mathbf{X}$, respectively. In particular, $\mathbf{I}$ represents the identity matrix. The $i$-th entry of a vector $\mathbf{x}\in \mathbb{C}^n$ is denoted by $[\mathbf{x}]_i$ for all $i\in \llbracket n \rrbracket$, where $\llbracket n \rrbracket$ represents the list $\{1,\cdots,n\}$. For any matrix $\mathbf{X}\in \mathbb{C}^{n_1 \times n_2}$, its $(i,j)$-th entry is denoted by $[\mathbf{X}]_{i,j}$ for all $i\in \llbracket n_1 \rrbracket$ and $j\in \llbracket n_2 \rrbracket$, and its $i$-th row is denoted by $[\mathbf{X}]_i\in \mathbb{C}^{1 \times  n_2}$ for all $i\in \llbracket n_1 \rrbracket$. Notations $ (\cdot)^T$, $(\cdot)^*$, $(\cdot)^{-1}$, $(\cdot)^\dag$, and $D(\cdot)$ represent the transpose, conjugate transpose, inverse, pseudo-inverse, and diagonal operators, respectively. Symbols $\otimes$ and $\circ$ denote the Kronecker and Hadamard products, respectively. We use $|(\cdot)|$, $\angle (\cdot)$, and $\mathrm{Re}(\cdot)$ to represent the magnitude, phase angle, and real part of a complex number, respectively. In addition, we use $\lfloor\cdot\rfloor$ to represent the floor operator, and use $\lfloor a \mod b \rceil $ to represent the modulo operation of $a$ with respect to $b$.
\section{Signal Model}
\label{SignalModel}
Consider a massive MIMO orthogonal frequency division multiplexing (OFDM) system, where one BS serves $K$ user equipments (UEs). Assume that the BS is equipped with a URA consisting of $M_x$ and $M_y$ array elements in the horizontal and vertical directions, respectively, while each UE has a single antenna. In the downlink communication, the signals received by all $K$ UEs on a certain subcarrier can be modeled as
\begin{equation}
\mathbf{y}\in\mathbb{C}^{K} =\mathbf{H}^*\mathbf{A}\mathbf{W}\mathbf{x} +\mathbf{n},
\label{receive}
\end{equation}
where $\mathbf{n}$ represents the noise. $\mathbf{x}\in \mathbb{C}^{K}$ denotes the transmit signal. $\mathbf{W}\in \mathbb{C}^{M \times K} = [\mathbf{w}_1, \ldots, \mathbf{w}_K]$ with $M=M_x M_y$ denotes the beamformer. $\mathbf{A}\in \mathbb{R}^{M \times M}=D(\mathbf{a})$ with $\mathbf{a}\in \mathbb{R}^M$ and $[\mathbf{a}]_i\ge 0$, $\forall i$, denotes the amplitude tapering matrix applied to the BS array for sidelobe suppression and channel capacity enhancement, as detailed in Section \ref{shaping}. In particular, $\mathbf{A}$ reduces to the identity matrix when no amplitude tapering is applied. $\mathbf{H}\in \mathbb{C}^{M \times K}=[\mathbf{h}_1, \ldots, \mathbf{h}_K]$ represents the multi-user spatial-frequency channel, with its $k$-th column being the channel between the BS and the $k$-th UE.
\par Assuming the presence of $L$ multipaths in the spatial-frequency channel $\mathbf{h}_k$, it can be modeled as
\begin{equation}
\mathbf{h}_{k}\in \mathbb{C}^{M} =\sum\limits_{l=1}^{L}{\alpha_{l,k}e^{-j2\pi f_s\tau_{l,k}}\mathbf{a}(\bm{\Theta}_{l,k})},
\label{r3}
\end{equation}
where $f_s$ denotes the frequency shift on a certain subcarrier in the OFDM system. The parameters for the $l$-th path in $\mathbf{h}_k$ comprise the complex gain $\alpha_{l,k}$, time delay $\tau_{l,k}$, elevation angle $\theta_{l,k}$, and azimuth angle $\varphi_{l,k}$. Note that in \eqref{r3}, the vector $\bm{\Theta}_{l,k}=[\Theta_{x,l,k},\Theta_{y,l,k}]^T$ with $\Theta_{x,l,k}=\sin(\theta_{l,k})\cos{(\varphi_{l,k})}$ and $\Theta_{y,l,k} = \sin(\theta_{l,k})\sin{(\varphi_{l,k})}$ indicates the direction of the $l$-th path in $\mathbf{h}_k$. In a narrowband scenario, the spatial steering vector $\mathbf{a}(\bm{\Theta}_{l,k})\in \mathbb{C}^{M}$ is formed by the Kronecker product between that in the horizontal direction, denoted as $\mathbf{a}_x(\Theta_{x,l,k})\in \mathbb{C}^{M_x}$, and that in the vertical direction, denoted as $\mathbf{a}_y(\Theta_{y,l,k})\in \mathbb{C}^{M_y}$\cite{tensor,b10}:
\begin{equation}
\left\{\begin{aligned}
& \mathbf{a}(\bm{\Theta}_{l,k}) = \mathbf{a}_x(\Theta_{x,l,k})\otimes\mathbf{a}_y(\Theta_{y,l,k}), \\
& \mathbf{a}_x(\Theta_{x,l,k})= [1,e^{-j\frac{2\pi}{\lambda} d\cdot \Theta_{x,l,k}},\cdots, e^{-j\frac{2\pi}{\lambda}(M_x-1)d\cdot \Theta_{x,l,k}}]^T,\\
& \mathbf{a}_y(\Theta_{y,l,k})=[1,e^{-j\frac{2\pi}{\lambda} d\cdot \Theta_{y,l,k}},\cdots, e^{-j\frac{2\pi}{\lambda}(M_y-1)d\cdot \Theta_{y,l,k}}]^T,
\end{aligned}\right.
\label{mw1}
\end{equation}
where $d$ is the inter-element spacing of a URA in both the horizontal and vertical directions, and $\lambda$ denotes the carrier wavelength.
\par The channel difference between URA and NUA configurations arises from variations in their spatial steering vectors. In the URA configuration, \eqref{mw1} can be rewritten as
\begin{equation}
\mathbf{a}(\bm{\Theta}_{l,k})=[e^{-j\frac{2\pi}{\lambda}\mathbf{s}_1^T{\cdot\bm{\Theta}}_{l,k}},\cdots,e^{-j\frac{2\pi}{\lambda}\mathbf{s}_{M}^T{\cdot\bm{\Theta}}_{l,k}}]^T,
\label{f2}
\end{equation}
where the vector $ \mathbf{s}_{m}=[s_{x,m}, \ \! s_{y,m}]^T$ with $s_{x,m}=\lfloor (m-1)/M_y \rfloor\cdot d$ and $s_{y,m}= \lfloor (m-1) \mod M_y \rceil \cdot d$ indicates the two-dimensional position of the $m$-th antenna, $\forall m \in \llbracket M \rrbracket$. However, the Kronecker product structure in the steering vector of URA, as defined in \eqref{mw1}, is absent in an NUA due to its irregular inter-element spacing. Hence, the steering vector in \eqref{f2} for a URA needs to be extended to accommodate an NUA, such that for all $m\in \llbracket M \rrbracket$,
\begin{equation}
\left[\mathbf{a}({\bm{\Theta}_{l,k}})\right]_m  = e^{-j\frac{2\pi}{\lambda}\mathbf{s}_m^T{\cdot\bm{\Theta}}_{l,k}}=e^{-j\frac{2\pi}{\lambda}(s_{x,m} \cdot\Theta_{x,l,k} +s_{y,m} \cdot\Theta_{y,l,k} )}, 
\label{beam2}
\end{equation}
where the position of the $m$-th antenna in both horizontal and vertical directions, denoted as $s_{x,m} \in \mathbb{R}$ and $s_{y,m} \in \mathbb{R}$, respectively, can be flexibly configured within the range of the array's aperture.
\par Note that for fairness, the inter-element spacing in an NUA, relative to the wavelength, is defined consistently with that in a URA. Specifically, for an NUA that possesses the same aperture size and the same number of antenna elements as a URA, its inter-element spacing is defined as $d$, which equals the inter-element spacing of the URA in both the horizontal and vertical directions, as shown in \eqref{mw1}. In other words, an NUA and a URA with identical aperture size and number of antenna elements are considered to have the same inter-element spacing.
\begin{figure*}[!t]
{\includegraphics[height=3.4cm]{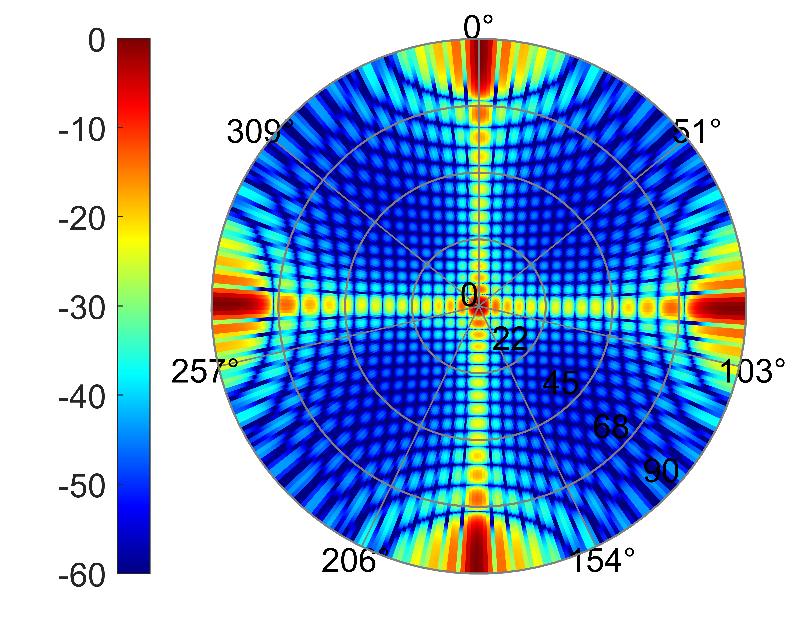}}
\hspace{-0.3cm}
{\includegraphics[height=3.4cm]{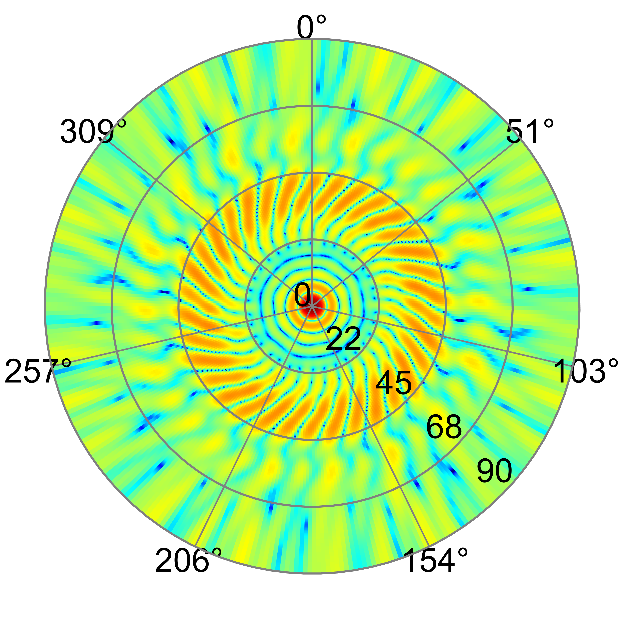}}
\hspace{-0.3cm}
{\includegraphics[height=3.4cm]{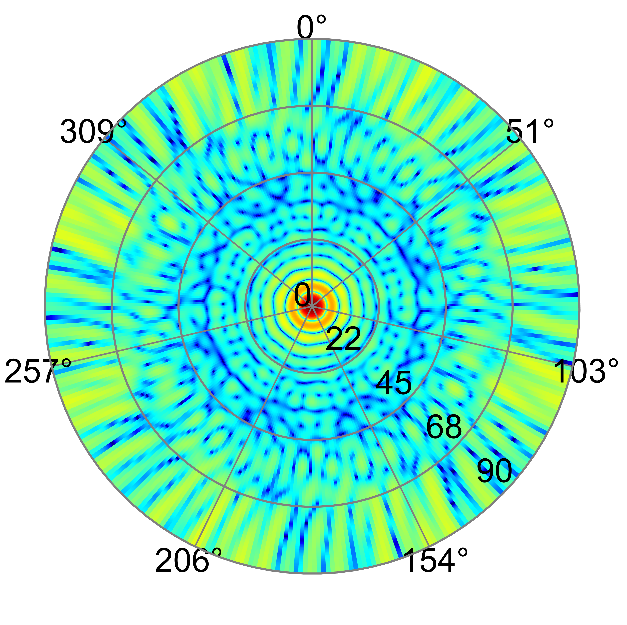}}
\hspace{-0.3cm}
{\includegraphics[height=3.4cm]{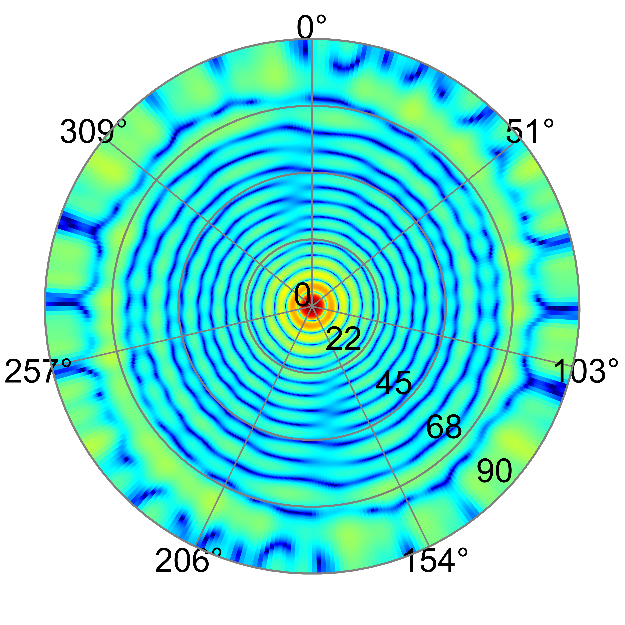}}
\hspace{-0.3cm}
{\includegraphics[height=3.4cm]{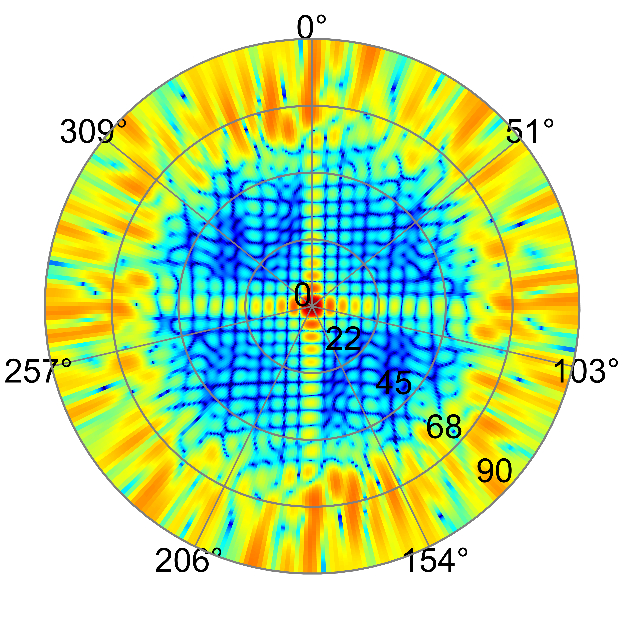}}
\hspace{-0.3cm}
\subfigure[URA]
{\includegraphics[height=3.4cm]{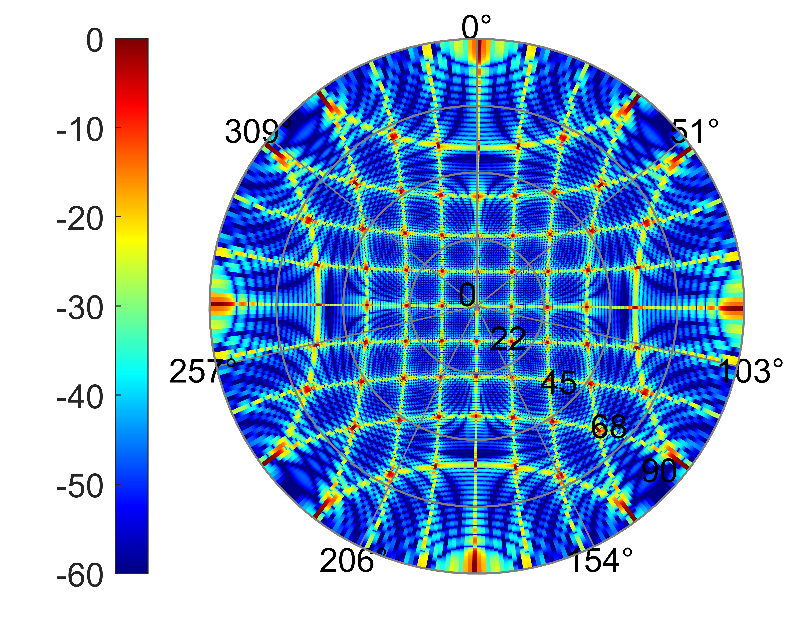}}
\hspace{-0.11cm}
\subfigure[GAM]
{\includegraphics[height=3.4cm]{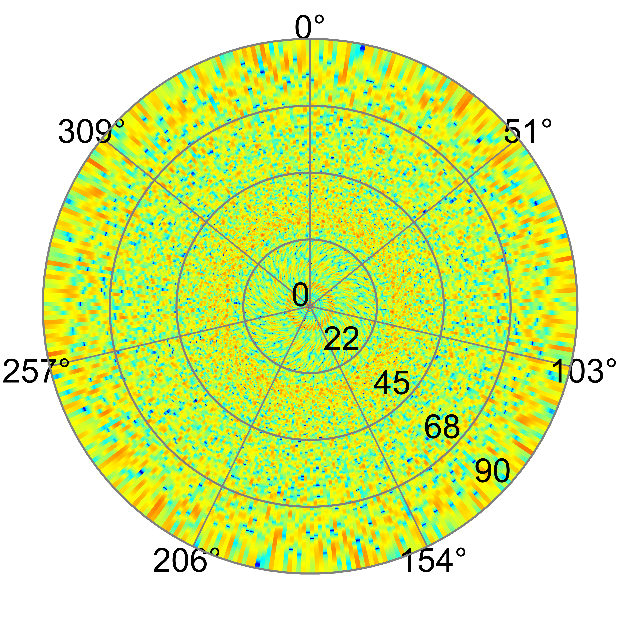}}
\hspace{-0.11cm}
\subfigure[DiscGAM]
{\includegraphics[height=3.4cm]{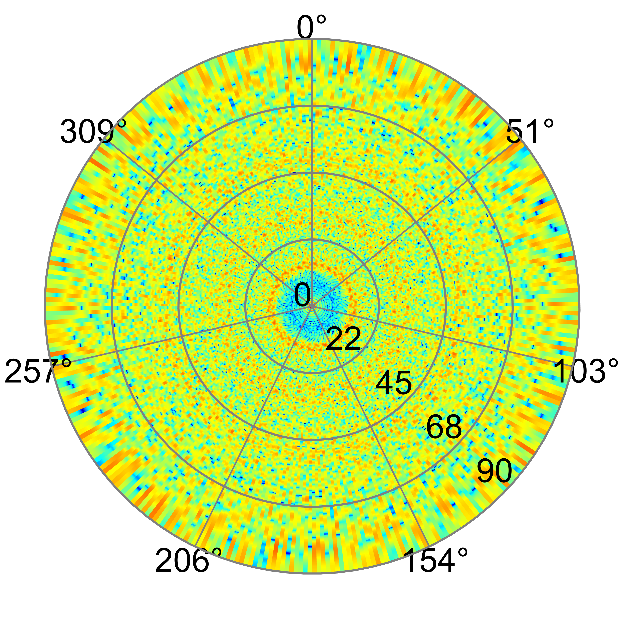}}
\hspace{-0.11cm}
\subfigure[SPM]
{\includegraphics[height=3.4cm]{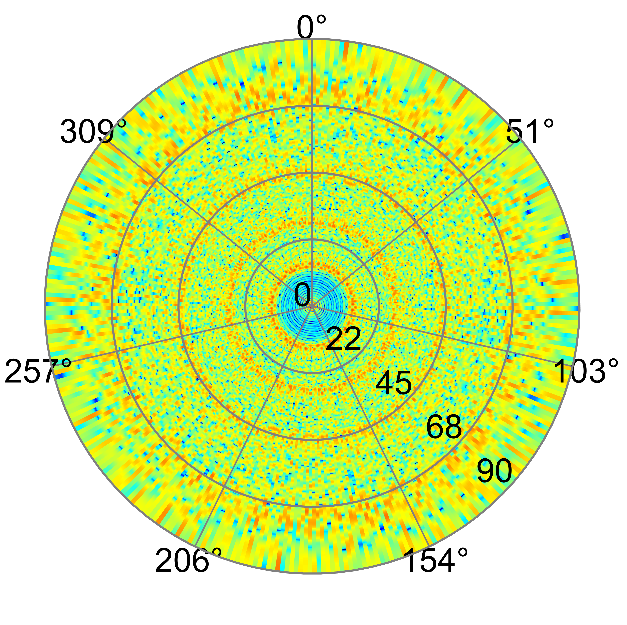}}
\hspace{-0.11cm}
\subfigure[HuD]
{\includegraphics[height=3.4cm]{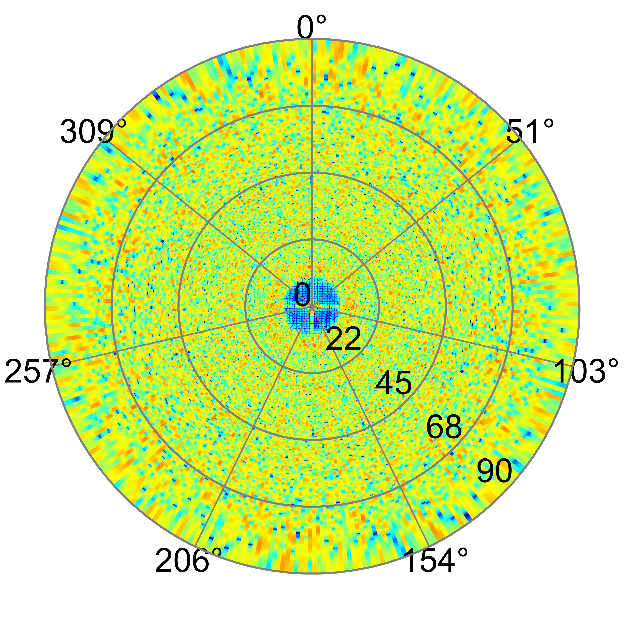}}
\caption{Comparison of the magnitude of the normalized array factor, as given in \eqref{e80}, for a URA in (a) and NUAs in (b)-(e), with the elevation angle $\theta_{l,k}$ varying over $[0^\circ,90^\circ]$ and the azimuth angle $\varphi_{l,k}$ over $[0^\circ,360^\circ]$. Results are shown in polar coordinates, where the radial and angular directions correspond to $\theta_{l,k}$ and $\varphi_{l,k}$, respectively. The URA and NUAs share the same aperture size and element number $M=225$, as illustrated in Fig. \ref{AllPattern}. The main lobe is steered towards $0^\circ $ in both elevation and azimuth. {\textbf{Rows}} 1 and 2 show the results when the inter-element spacings are set to $d=\epsilon \lambda$ with constants $\epsilon=1 \ {\text{and}} \ 5$, respectively.}
\label{figureNet111}
\end{figure*}
\begin{figure*}[!t]
\hspace{-0.15cm}
{\includegraphics[height=3.15cm]{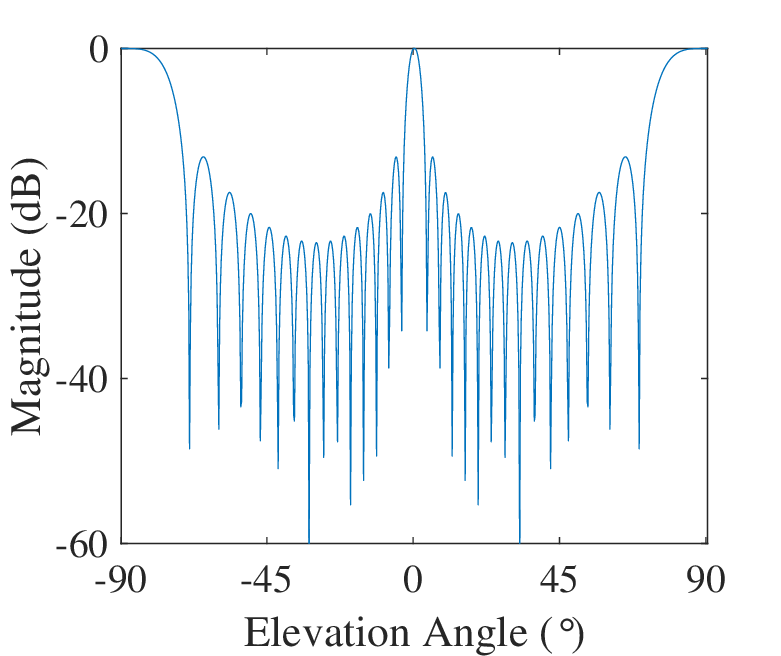}}
\vspace{-0.0cm}
\hspace{-0.5cm}
{\includegraphics[height=3.15cm]{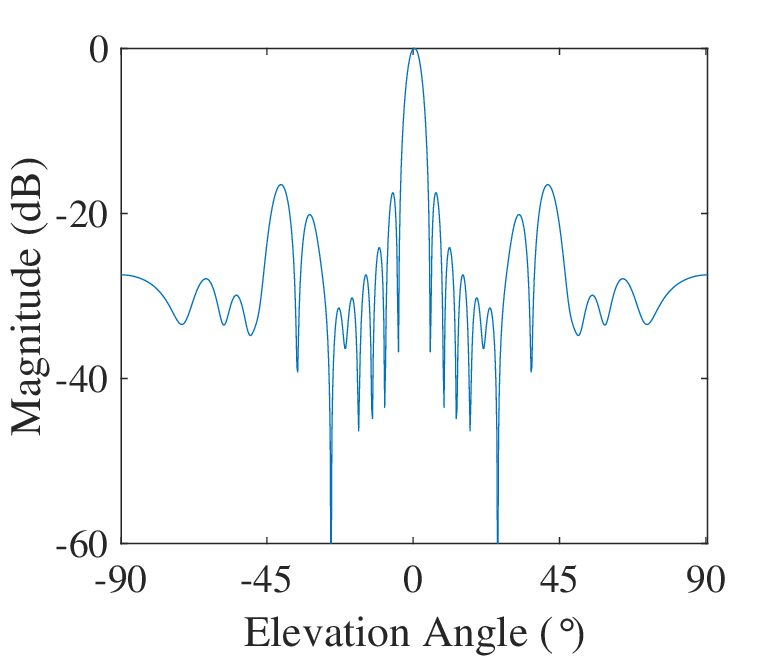}}
\hspace{-0.0cm}
\hspace{-0.5cm}
{\includegraphics[height=3.15cm]{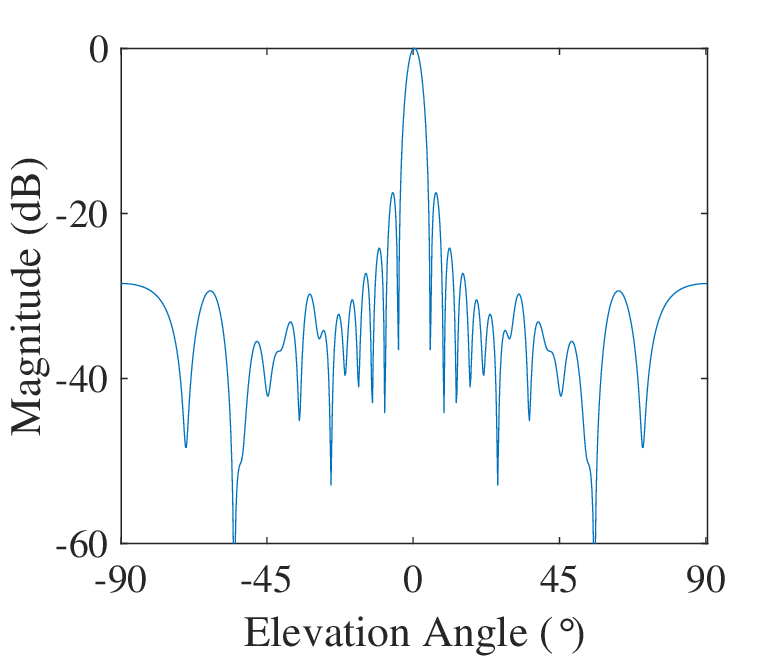}}
\hspace{-0.0cm}
\hspace{-0.5cm}
{\includegraphics[height=3.15cm]{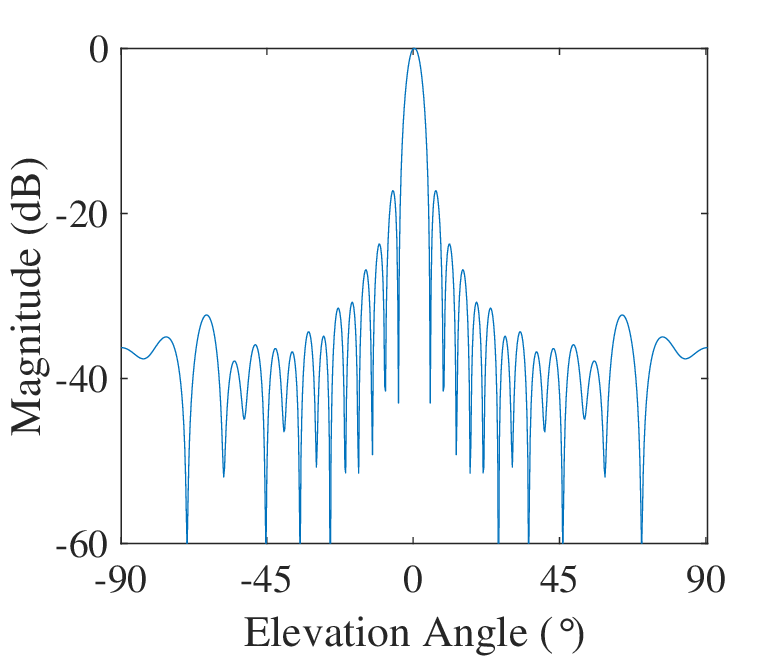}}
\hspace{-0.0cm}
\hspace{-0.5cm}
{\includegraphics[height=3.15cm]{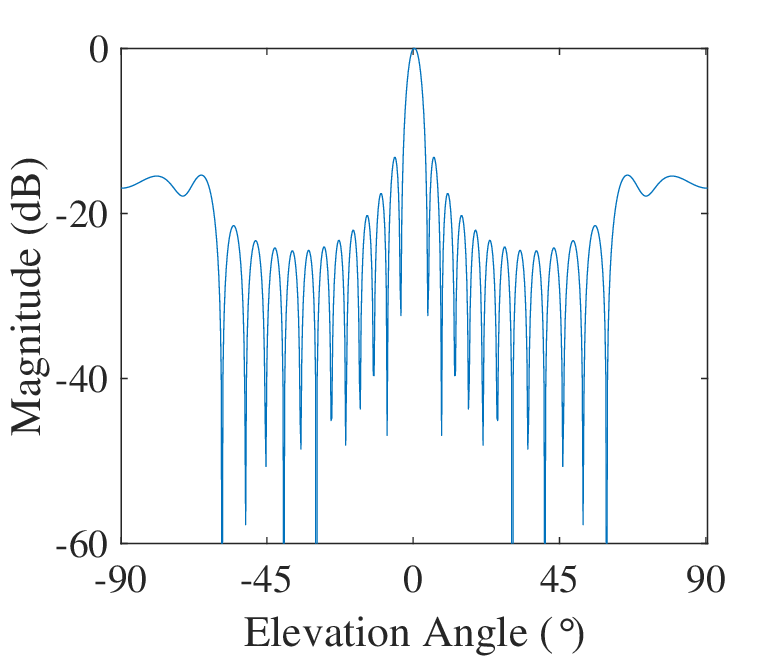}}
\hspace{-0.0cm}
\hspace{-0.5cm}
\subfigure[URA]
{\includegraphics[height=3.15cm]{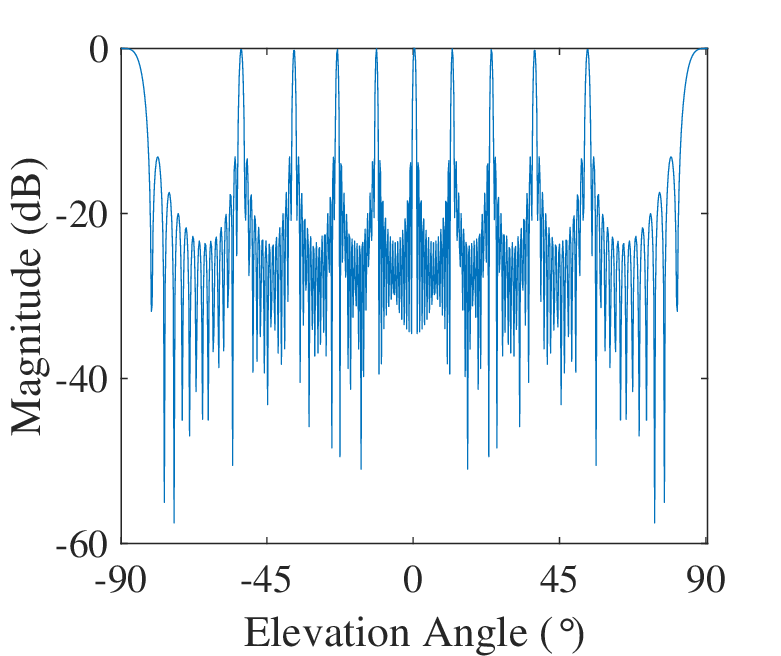}}
\vspace{-0.0cm}
\hspace{-0.5cm}
\subfigure[GAM]
{\includegraphics[height=3.15cm]{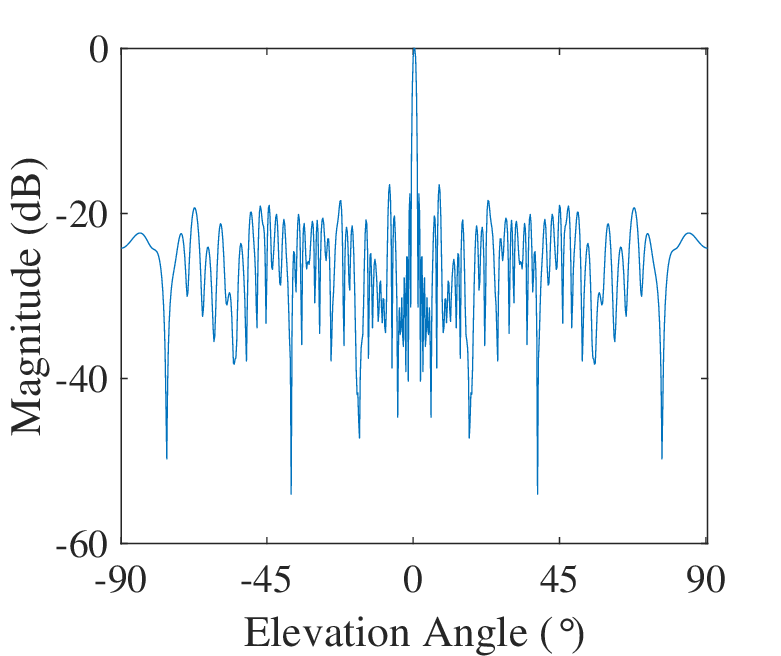}}
\hspace{-0.0cm}
\hspace{-0.5cm}
\subfigure[DiscGAM]
{\includegraphics[height=3.15cm]{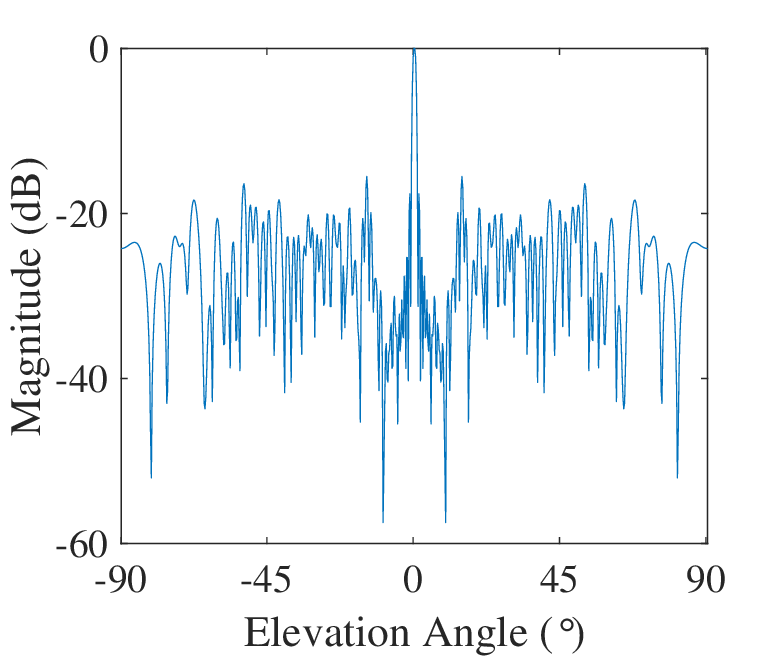}}
\hspace{-0.0cm}
\hspace{-0.5cm}
\subfigure[SPM]
{\includegraphics[height=3.15cm]{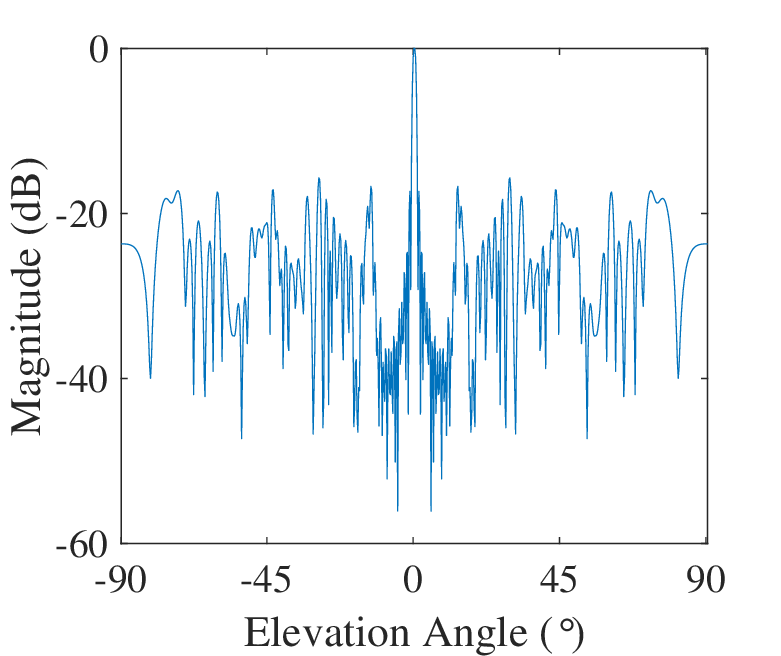}}
\hspace{-0.0cm}
\hspace{-0.5cm}
\subfigure[HuD]
{\includegraphics[height=3.15cm]{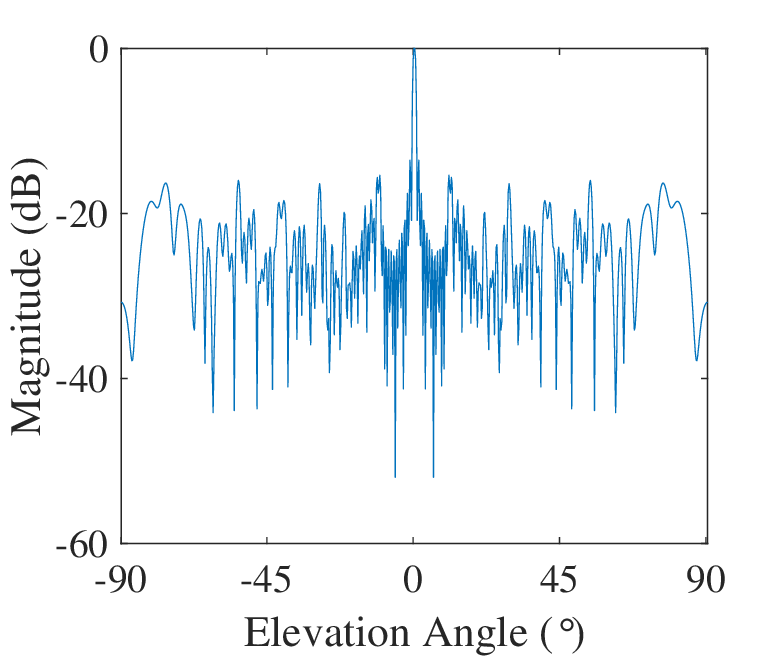}}
\hspace{-0.0cm}
\hspace{-0.3cm}
\caption{Comparison of the magnitude of the normalized array factor, as given in \eqref{e80}, for a URA in (a) and NUAs in (b)-(e), with the azimuth angle fixed at $\varphi_{l,k}=0^\circ $ and the elevation angle $\theta_{l,k}$ varying over $[-90^\circ,90^\circ]$. All array configurations follow those in Fig. \ref{figureNet111}. The main lobe is steered towards $0^\circ $ in elevation. {\textbf{Rows}} 1 and 2 show the results when the inter-element spacings are set to $d=\epsilon \lambda$ with constants $\epsilon=1 \ {\text{and}} \ 5$, respectively.}
\label{figureNet1}
\end{figure*}
\section{Advantages of Non-Uniform Arrays}
\label{advan}
This section presents the advantages of the proposed NUAs by evaluating the radiation patterns and favorable propagation. Specifically, the design of NUA element distributions draws inspiration from mutual-information-optimal constellation shaping in the signal domain\cite{GAM,GAM2,spiral}. In constellation design, aperiodic geometric arrangements such as golden angle-based and spiral-based patterns\cite{GAM,GAM2,spiral} are known to achieve favorable geometric properties and enhanced mutual information compared with conventional uniform grids. By analogy, we apply these well-structured aperiodic patterns to the spatial domain as antenna element distributions, leveraging their inherent geometric advantages for grating lobe suppression and improved channel characteristics. Following this philosophy, we consider the GAM, DiscGAM, SPM, and HuD distributions as the basic NUA pattern designs, as shown in Fig. \ref{AllPattern}.
\subsection{Grating Lobe-Free Effect}
Based on \eqref{f2} and \eqref{beam2}, the magnitude of the normalized array factor for an arbitrary array distribution is given by \cite{HUDX}:
\begin{equation}
|A(\bm{\Theta}_{l,k})|= \frac{1}{M} \left|\sum\nolimits_{m=1}^{M} e^{-j\frac{2\pi}{\lambda}\mathbf{s}_{m}^T{\cdot\bm{\Theta}}_{l,k}}\right|,
\label{e80}
\end{equation}
where $\bm{\Theta}_{l,k}\in \mathbb{R}^2=[\sin(\theta_{l,k})\cos{(\varphi_{l,k})},\, \sin(\theta_{l,k})\sin{(\varphi_{l,k})}]^T$ is given in \eqref{r3}. According to the Nyquist-Shannon sampling theorem\cite{Shannon1,Shannon3}, the spatial sampling interval (i.e., inter-element spacing) in a uniformly distributed array should be smaller than half a wavelength to avoid spatial aliasing artifacts. Conversely, a larger inter-element spacing generally leads to the emergence of more grating lobes\cite{GL1,GL2}. For example, Figs. \ref{figureNet111} and \ref{figureNet1} depict the magnitude of the normalized array factor in \eqref{e80}, with inter-element spacing set to $d=\epsilon \lambda$, where $\lambda = 6$ cm and $\epsilon$ is a constant equal to $1$ or $5$. As illustrated in Figs. \ref{figureNet111}(a) and \ref{figureNet1}(a), grating lobes appear in the URA when $\epsilon=1$ and become more conspicuous as the inter-element spacing increases to $\epsilon=5$. In contrast, NUAs with well-designed element distributions, such as GAM, DiscGAM, SPM, or HuD distribution, remain free of grating lobes for both $\epsilon=1$ and $\epsilon=5$.
\par
\begin{figure}[tbp]
\centerline{\includegraphics[width=7.1cm]{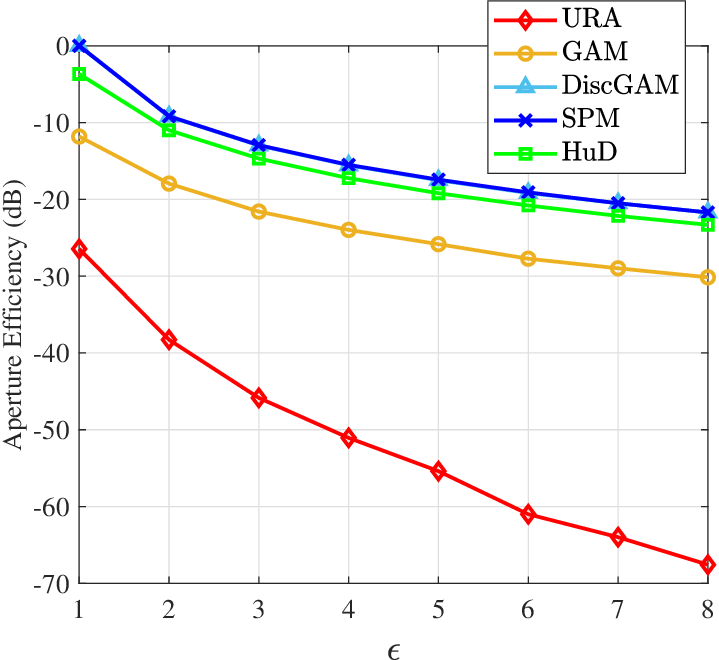}}
\centering
\caption{AE versus inter-element spacing $d=\epsilon\lambda$ with the constant $\epsilon$ for a URA and NUAs with GAM, DiscGAM, SPM, and HuD distributions.}
\label{AER}
\end{figure}

 The presence of grating lobes in the URA inevitably reduces the antenna aperture efficiency (AE). Based on \eqref{r3}, AE in the direction of the $l$-th path of the $k$-th UE's channel can be measured as \cite{GL1,GL2}:
\begin{equation}
\mathrm{AE}=\frac{|A(\bm{\Theta}_{l,k})|^2\cdot \theta_{\text{exc}}}{|A(\bm{\Theta}_{l,k})|^2\! \cdot \! \frac{\pi}{2} +\sum^J_{j=1}|A(\hat{\bm{\Theta}}_{l,k,j})|^2 \!\cdot\! \frac{\cos(\theta_{l,k})}{\max\{\cos (\hat{\theta}_{l,k,j}),\, \eta\}}},
\label{e9}
\end{equation}
where $\bm{\Theta}_{l,k}$ and $\hat{\bm{\Theta}}_{l,k,j}=[\sin(\hat{\theta}_{l,k,j})\cos{(\hat{\varphi}_{l,k,j})},\ \sin(\hat{\theta}_{l,k})\cdot$ $\sin{(\hat{\varphi}_{l,k,j})}]^T$ denote the direction vectors of the main lobe and the $j$-th grating lobe, respectively. Owing to the symmetry of the array factor, the number of grating lobes, denoted as $J$, is counted in the half-space where $\hat{\theta}_{l,k}\in [0,\pi/2]$ and $\hat{\varphi}_{l,k}\in [0,\pi)$. A non-zero constant $\eta=0.1$ is used to avoid division by zero when the $j$-th grating lobe emerges at the azimuth angle $\hat{\theta}_{l,k,j}=\pi/2$, $\forall j\in \llbracket J \rrbracket$. The angle $\theta_{\text{exc}}\in (0,\pi/2]$ represents the radius of the circular exclusion region in the array's radiation pattern\cite{HUDX}. Following the experimental setup
in Fig. \ref{figureNet1}, we evaluate the AE score versus inter-element spacing in Fig. \ref{AER}. As the inter-element spacing increases, the AE of the URA rapidly decreases due to the influence of grating lobes on antenna directivity. Conversely, all tested NUAs in Fig. \ref{AER} exhibit more stable AE scores owing to their ability to mitigate grating lobes.
\subsection{Favorable Propagation}
\label{FavPro}
We investigate the channel characteristics of massive MIMO with NUAs and compare them to those of conventional URAs. In multi-user scenarios, the grating lobe suppression will reduce inter-user interference, thereby enhancing the orthogonality of the multi-user channel $\mathbf{H}$, as shown in \eqref{receive}. Consider the channel covariance matrix (CCM), i.e., $\mathbf{H}^* \mathbf{H}$. Under the URA configuration with a large inter-element spacing, the off-diagonal elements in CCM cannot be effectively suppressed due to the emergence of grating lobes. In contrast, for a given number of antenna elements and the same size of array aperture, NUAs can effectively avoid grating lobes, leading to a more diagonalizable CCM. 
\par The diagonalizable CCM benefits the multi-user channel of massive MIMO systems from favorable propagation. The favorable propagation property characterizes the channel orthogonality and emerges in MIMO systems when the BS antenna number is sufficiently large such that
\begin{equation}
  \mathbf{h}_i^* \mathbf{h}_j/M \rightarrow 0, \ \forall i\neq j, \ \text{as} \ M \rightarrow \infty.
  \label{as1}
\end{equation}
Notably, channel capacity reaches its maximum when multi-user channels are completely orthogonal among different UEs. Specifically, under the additive white Gaussian noise assumption that $\mathbf{n} \sim \mathcal{C} \mathcal{N}(0, \sigma^2 \mathbf{I})$ in \eqref{receive}, the achievable sum rate with respect to the channel $\mathbf{H}$ under the maximal ratio combining (MRC) criterion, denoted as $C_1(\mathbf{H})$, is defined as
\begin{equation}
C_1(\mathbf{H}) \triangleq \sum\nolimits_{k=1}^K\log_2(1+\gamma_k),
\label{cpc0}
\end{equation}
where $\gamma_k$ denotes the signal-to-interference-plus-noise ratio of the $k$-th UE, defined as
\begin{equation}
\gamma_k=\frac{\big| \mathbf{w}_k^* \cdot \mathbf{h}_k\big|^2}{\sum_{i \neq k}^M\big|\mathbf{w}^*_i \cdot \mathbf{h}_k \big|^2+\sigma^2}.
\label{SINR}
\end{equation}
Assume that the zero-forcing (ZF) beamformer $\mathbf{W}=[\mathbf{w}_1, \ldots, \mathbf{w}_K]$ is employed\cite{ZF}, with unit power allocated to each UE. Then, $C_1(\mathbf{H})$ is upper-bounded by the maximum sum rate with respect to the channel $\mathbf{H}$, denoted as $C_2(\mathbf{H})$, such that
\begin{equation}
C_1(\mathbf{H})\le C_2(\mathbf{H}) \triangleq
\log_2 \det \left(
\mathbf{I} + \frac{1}{\sigma^2} \mathbf{H}\mathbf{H}^* \right).
\label{cpc}
\end{equation}
The equality $C_1(\mathbf{H})=C_2(\mathbf{H})$ holds when the channel vectors of different UEs are mutually orthogonal, i.e., $\mathbf{h}_i^* \mathbf{h}_j = 0, \forall i\neq j$. Conversely, $C_1(\mathbf{H})<C_2(\mathbf{H})$ if these channel vectors are not completely orthogonal, i.e., there exists at least one pair $i\neq   j$ such that $\mathbf{h}_i^* \mathbf{h}_j \neq 0$.
\section{Shaping the Non-Uniform Arrays}
\label{shaping}
\begin{figure}[!t]
\subfigure[\hspace{-0.5cm}]
{\includegraphics[width=4.3cm]{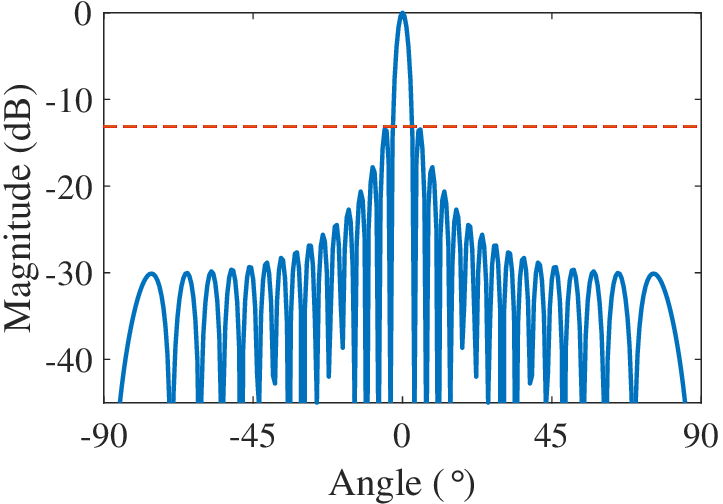}}
\vspace{-0.0cm}
\hspace{-0.0cm}
\subfigure[\hspace{-0.55cm}]
{\includegraphics[width=4.3cm]{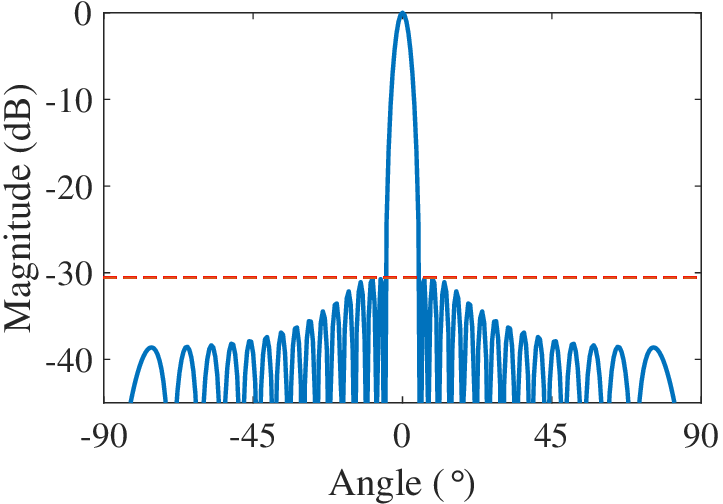}}
\hspace{-3cm}
\caption{The normalized array factors without and with amplitude tapering are shown in (a) and (b), respectively. Specifically, a uniform linear array with $N=32$ and half-wavelength spacing is considered. A Taylor window-based amplitude tapering with a sidelobe level of $-30$ dB is applied.}
\label{AF_COM}
\end{figure}

Traditional array design is typically guided by geometric or field‑based criteria, such as main lobe width, sidelobe level, or grating lobe suppression\cite{HUD,HUD2,HUD3,HUDX,HUD0,sate1,FP3}. Although these metrics characterise the radiation pattern, they only indirectly relate to the fundamental goal of a communication system: to convey information with minimal distortion. From the communication perspective, the array response vector is inherently determined by both the geometric distribution of antenna elements and their excitation amplitudes. Optimizing the geometric distribution of antenna elements is referred to as geometric shaping, whereas optimizing the excitation amplitudes is known as amplitude tapering. Specifically, amplitude tapering reshapes the array factor for sidelobe suppression, as illustrated in Fig. \ref{AF_COM}, which is achieved by adjusting the excitation weights across antenna elements under a certain power budget. By shaping the amplitude weights and antenna positions either independently or jointly, NUAs can be tailored to more favorable characteristics of MIMO channels.
\par It is noteworthy that the array shaping can be interpreted as a spatial analogue of constellation shaping by treating each antenna element as a spatial constellation point\cite{GAM,GAM2,spiral}. Under this interpretation, the antenna positions correspond to geometric shaping, while the excitation amplitudes correspond to probability shaping, both aiming to maximize the mutual information\cite{Shannon1,Shannon3}. Consequently, we refer to the proposed array shaping framework as an EMIT-based design approach. Our second key insight is that the beam pattern in the angular domain is the Fourier transform of the spatial sampling of the antenna elements. Since the Fourier transform is an invertible operation, it does not reduce mutual information; any degradation in information transfer must therefore come from the spatial sampling pattern itself rather than from the transformation to the angular domain. This observation allows us to treat different array geometries as different spatial sampling operators and to evaluate them rigorously through the capacity of the resulting MIMO channel.
\subsection{Geometric Shaping}
\label{GSsec}
Assume that the BS is equipped with an array that has a fixed number of elements and a fixed aperture size, while allowing variable element geometries within the aperture. Under this assumption, we treat the array geometric shaping problem as an antenna selection problem\cite{AS2,AS1}, which aims to select $M$ antenna elements from an antenna-dense URA, consisting of $TM_x$ and $TM_y$ antenna elements in the horizontal and vertical directions, respectively, with $M=M_xM_y$ and $T\gg 1$. The criterion of antenna selection is to maximize the capacity of the multi-user MIMO channel $C_2(\mathbf{H})$, as defined in \eqref{cpc}, where $\mathbf{H}\in \mathbb{C}^{M \times K}=[\mathbf{h}_1, \ldots, \mathbf{h}_K]$ is modeled in \eqref{r3}-\eqref{beam2} and varies with the positions of $M$ selected antennas. It is worth mentioning that we choose the capacity $C_2(\mathbf{H})$ rather than $C_1(\mathbf{H})$ to be maximized. This is because the value of $C_1(\mathbf{H})$ depends not only on the channel $\mathbf{H}$ but also on the beamformer $\mathbf{W}=[\mathbf{w}_1, \ldots, \mathbf{w}_K]$. Hence, maximizing $C_1(\mathbf{H})$ requires a joint optimization of $\mathbf{H}$ and $\mathbf{W}$, which is intractable and more computationally expensive than optimizing only $\mathbf{H}$ via maximizing $C_2(\mathbf{H})$. Although maximizing $C_1(\mathbf{H})$ and $C_2(\mathbf{H})$ may yield different results, they have a close relationship since $C_1(\mathbf{H})$ is upper-bounded by $C_2(\mathbf{H})$, as shown in \eqref{cpc}. Using the proposed geometric shaping algorithm, the bound $C_2(\mathbf{H})=C_1(\mathbf{H})$ is approximately reachable in some scenarios, as shown in the experimental results in Figs. \ref{CapComp1} and \ref{AVE}.
\par Due to the use of antenna-dense URA with $T\gg 1$, selecting all possible combinations of $M$ out of $T^2M$ antennas leads to the curse of dimensionality issue \cite{AS1,AS2}. This issue renders traditional antenna selection approaches based on convex optimization \cite{AS1} computationally infeasible, especially when handling the two-dimensional array scenario. In contrast, greedy-based antenna selection approaches are computationally efficient and are thus investigated in this paper. However, traditional greedy-based algorithms \cite{AS2} perform poorly in the geometric shaping problem where $T\gg 1$, since they are non-convex and sensitive to the initialization. To address this issue, we propose a greedy-based antenna position update (APU) algorithm for geometric shaping. APU involves up to $N$ iterations, as summarized in Algorithm \ref{GS}. For the algorithm initialization, we construct the selected antenna set as an $M$-element NUA with the HuD (or alternatively GAM, DiscGAM, or SPM) distribution, and construct the candidate antenna set as a URA consisting of $T^2M$ elements.
\par
\begin{figure}[!t]
\centerline{\includegraphics[width=8.9cm]{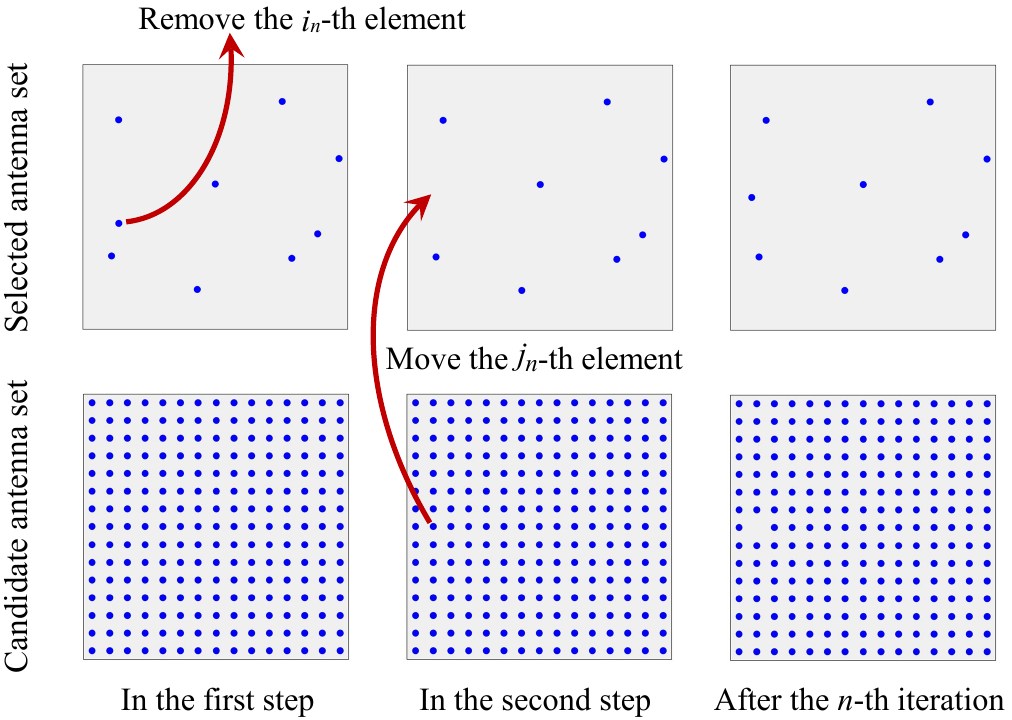}}
\centering
\caption{Illustration of element updates for the selected antenna set, consisting of $M=9$ elements, and for the candidate antenna set, consisting of $(T^2M-n)$ elements with $T=5$, in the $n$-th iteration of APU with $n=0$. In the first step of the $n$-th iteration, the $i_n$-th element in the selected antenna set is removed. In the second step, the $j_n$-th element in the candidate antenna set is moved to the selected antenna set.}
\label{antennaupdate}
\end{figure}

 As illustrated in Fig. \ref{antennaupdate}, the $n$-th iteration of APU involves two steps, before which the selected antenna set and candidate antenna set contain $M$ and $(T^2M-n)$ elements, respectively. In the first step of the $n$-th iteration, the selected antenna set removes one element, thus leaving $(M-1)$ elements remaining. In the second step of the $n$-th iteration, one element in the candidate antenna set is moved to the selected antenna set, so that the selected antenna set and candidate antenna set contain $M$ and $(T^2M-n-1)$ elements, respectively, after the second step. In this way, the selected antenna set achieves the update of one antenna element through the $n$-th iteration of APU. Before providing optimization problems for the $n$-th iteration of APU, let us denote the multi-user channels corresponding to the selected antenna set and candidate antenna set in the $n$-th iteration as $\mathbf{H}_n\in \mathbb{C}^{M\times K}$ and $\mathbf{P}_n\in \mathbb{C}^{(T^2M-n)\times K}$, respectively. Based on the a priori known antenna positions and channel state information, $\mathbf{H}_n$ and $\mathbf{P}_n=[\mathbf{p}_{n,1}, \ldots, \mathbf{p}_{n,K}]$ can be modeled following \eqref{r3}-\eqref{f2}. Specifically, $\mathbf{p}_{n,k}\in \mathbb{C}^{(T^2M-n)}$ is modeled similarly to $\mathbf{h}_k$ in \eqref{r3}, with $\mathbf{a}(\bm{\Theta}_{l,k})$ in \eqref{f2} replaced by $\tilde{\mathbf{a}}(\bm{\Theta}_{l,k})\in \mathbb{C}^{(T^2M-n)}\!=\![e^{-j\frac{2\pi}{\lambda}\tilde{\mathbf{s}}_1^T{\cdot\bm{\Theta}}_{l,k}},\cdots,e^{-j\frac{2\pi}{\lambda}\tilde{\mathbf{s}}_{(T^2M-n)}^T{\cdot\bm{\Theta}}_{l,k}}]^T$, where $ \tilde{\mathbf{s}}_j=[\tilde{s}_{x,j}, \tilde{s}_{y,j}]^T$ denotes the two-dimensional position of the $j$-th element in the candidate antenna set, $\forall j\in \llbracket T^2M -n\rrbracket$. Based on these channel models, the first and second steps in the $n$-th iteration of APU yield optimization problems in \eqref{q2one} and \eqref{q2two}, respectively, as
\begin{numcases}{}
i_n= \mathop {\arg \max }\limits_{i\in \llbracket M \rrbracket} C_2\Big(\mathbf{H}_n^{(i)}\Big),
\label{q2one}
\\
j_n= \mathop {\arg \max }\limits_{j\in \llbracket T^2M -n \rrbracket}  C_2\Big(\mathbf{H}_n^{(i_n,j)}\Big),
\label{q2two}
\end{numcases}
where $\mathbf{H}_n^{(i)}\in \mathbb{C}^{(M-1)\times K}$ is formed by deleting the $i$-th row of $\mathbf{H}_n$, and  $\mathbf{H}_n^{(i_n,j)}\in\mathbb{C}^{M\times K}=\big[(\mathbf{H}_n^{(i_n)})^T,[\mathbf{P}_n]^T_j\big]^T$ is formed by moving the $j$-th row of $\mathbf{P}_n$ to the last row of $\mathbf{H}_n^{(i_n)}$. After solving \eqref{q2one} and \eqref{q2two}, Algorithm \ref{alg1} updates the channels $\mathbf{H}_{n+1}$ and $\mathbf{P}_{n+1}$, as shown in Line 4, used for the next iteration.
\par It is noteworthy that directly solving \eqref{q2two} is computationally expensive due to the enormous elements within the candidate antenna set in scenarios where $T\gg 1$. To enhance computational efficiency, we apply the Sherman-Morrison formula for determinants
\cite{keri2001sherman,AS2} to the objective of \eqref{q2two}, yielding
{\small{\begin{equation}
C_2\Big(\mathbf{H}_n^{(i_n,j)}\Big)=C_2\Big(\mathbf{H}_n^{(i_n)}\Big)+\log _2\Big(1+ \frac{1}{\sigma^2}[\mathbf{P}_n]_j\cdot f\big(\mathbf{H}_n^{(i_n)}\big)\cdot [\mathbf{P}_n]^*_j \Big),
\label{q2four}
\end{equation}}}where $f(\mathbf{H}_n^{(i_n)}) \triangleq\big(\mathbf{I}+ (\mathbf{H}_n^{(i_n)})^* \mathbf{H}_n^{(i_n)}/\sigma^2\big)^{-1}$. Based on \eqref{q2four}, \eqref{q2two} can be equivalently reformulated as
\begin{equation}
j_n= \mathop {\arg \max }\limits_{j\in \llbracket T^2M -n \rrbracket} \  [\mathbf{P}_n]_j\cdot f\big(\mathbf{H}_n^{(i_n)}\big)\cdot [\mathbf{P}_n]^*_j,
\label{equal2}
\end{equation}
which can be solved within polynomial time.
\subsection{Amplitude Tapering}
In addition to the geometric shaping introduced in Section \ref{GSsec}, amplitude tapering can be further applied to the BS array to enhance  channel capacity. Note that the channel capacities defined in \eqref{cpc0} and \eqref{cpc} do not involve amplitude tapering for the BS array and correspond to a signal model in \eqref{receive} where the amplitude tapering matrix $\mathbf{A}=D(\mathbf{a})$ reduces to the identity matrix. In contrast, when taking amplitude tapering into account, the capacity $C_2(\mathbf{H})$ in \eqref{cpc} is extended to 
\begin{equation}
C_3(\mathbf{H},\mathbf{a}) \triangleq
\log_2 \det \left(
\mathbf{I} + \frac{1}{\sigma^2} D(\mathbf{a}) \mathbf{H}\mathbf{H}^* D(\mathbf{a})
\right),
\label{cap3}
\end{equation}
which is dependent not only on the channel $\mathbf{H}$ but also on the amplitude tapering vector $\mathbf{a}\in \mathbb{C}^{M}$ as excitation weights over $M$ elements of the BS array. The equality $C_3(\mathbf{H},\mathbf{a}) = C_2(\mathbf{H})$ holds when no amplitude tapering is applied, i.e., $\mathbf{A}=\mathbf{I}$, while an appropriate amplitude tapering has the potential to achieve $C_3(\mathbf{H},\mathbf{a})>C_2(\mathbf{H})$. 
\par To avoid trivial unbounded solutions, practical constraints need to be imposed on the amplitude tapering. Specifically, we consider the joint power budget and non-negativity constraints, yielding the following optimization problem:
\begin{equation}
\label{objAT}
\max_{\mathbf{a}} \, C_3(\mathbf{H},\mathbf{a}) \quad 
\text{s.t.} \ \|\mathbf{a}\|^2_2 \le P, \ [\mathbf{a}]_m \ge 0, \ \forall m\in \llbracket M \rrbracket,
\end{equation}
where $P$ represents the power budget. Problem \eqref{objAT} is non-convex due to the quadratic dependence of $D(\mathbf{a})\mathbf{H}\mathbf{H}^*D(\mathbf{a})$ on $\mathbf{a}$ in the objective function. Nevertheless, this objective function is smooth and differentiable with respect to  $\mathbf a$, and thus \eqref{objAT} can be effectively solved using the projected gradient ascent (PGA) method\cite{bertsekas1997nonlinear}, as summarized in Algorithm \ref{algpga_taper}. 
\par The $n$-th iteration of PGA updates the variable $\mathbf{a}_{n+1}$ as
\begin{equation}
 \mathbf{a}_{n+1} = \Pi_{\mathcal{F}}(\mathbf{a}_n + \alpha\,\nabla_{\mathbf{a}_n} C_3(\mathbf{H},\mathbf{a}_n)).
\end{equation}
The gradient of $C_3(\mathbf{H},\mathbf{a}_n)$ with respect to $\mathbf a_n$ can be calculated as
\begin{equation}
[\nabla_{\mathbf{a}_n} C_3(\mathbf{H},\mathbf{a}_n )]_i
\!= \!\frac{2\, \mathrm{Re}\! 
\left([(\mathbf{I} \!+\! \mathbf{A}_n\mathbf{R} \mathbf{A}_n/ {\sigma^2})^{-1}\,\mathbf{A}_n\mathbf{R}]_{i,i}
\right)}{\sigma^2\ln 2},\forall i,
\end{equation}
where $\mathbf{A}_n=D(\mathbf{a}_n)$ and $\mathbf{R} = \mathbf{H}\mathbf{H}^*$. The step size $\alpha> 0 $ is selected through a backtracking line search based on two parameters $\beta<1$ and $c<1$, as detailed in Lines 6-10 of Algorithm \ref{algpga_taper}. For any $\mathbf{x}\in \mathbb{C}^M$, the operator $\Pi_{\mathcal{F}} (\mathbf{x})$ projects it onto the feasible set $\mathcal{F}$, defined as
\begin{equation}
\mathcal{F} \triangleq
\left\{
\mathbf{x} \,\big|\,
\|\mathbf{x}\|^2_2 \le P \ {\text{and}} \ 
[\mathbf{x}]_m \ge 0, \forall m\in \llbracket M \rrbracket
\right\},
\end{equation}
and has a closed-form expression as
\begin{equation}
\! \Pi_{\mathcal{F}}({\mathbf{x}})= \begin{cases}\mathbf{x}, & \begin{array}{l}
\!\!\!\!\!\! \text {if } \mathbf{x} \in \mathcal{F}, 
\end{array} \\
{\sqrt{P}\cdot \max(\mathbf{0},\mathbf{x})}\big/{\|\max(\mathbf{0},\mathbf{x})\|_2}, & \!\!\!\! \text {\,otherwise.}\end{cases}
\end{equation}
\par The amplitude tapering and geometric shaping techniques can be naturally combined in massive MIMO systems, where the antenna geometry is first optimized via Algorithm \ref{GS}, and then the excitation amplitudes of antennas are optimized via Algorithm \ref{algpga_taper} to enhance the channel capacity. This combined design is referred to as joint geometric shaping and amplitude tapering.
\section{Extension to Wideband Scenarios}
\label{beamsquint}
In massive MIMO systems with a large number of antennas, considerable time delays occur across various array elements for identically transmitted data symbols, leading to the spatial-wideband effect\cite{bl}. This effect is inherent to large-scale arrays. Moreover, the spatial-wideband effect in OFDM systems will expand to encompass both spatial and frequency domains, leading to the beam squint phenomenon\cite{bl2,gl}. This is due to the fact that the spatial steering vector becomes highly dependent on the operating frequency, necessitating a modification of $\mathbf{a}(\bm{\Theta}_{l,k})$, as stated in \eqref{f2}, into\cite{bl,gl,bl2}
\begin{equation}
\mathbf{a}(\bm{\Phi}_{l,k,q})=[e^{-j\frac{2\pi}{\lambda} \mathbf{s}_1^T\cdot {\bm{\Phi}}_{l,k,q}},\cdots,e^{-j\frac{2\pi}{\lambda} \mathbf{s}_{M}^T\cdot{\bm{\Phi}}_{l,k,q}}]^T,
\label{g3}
\end{equation}
where ${\bm{\Phi}}_{l,k,q} = (1+\frac{f_q}{f_c}) \bm{\Theta}_{l,k}$ indicates the direction of the $l$-th path of the $k$-th UE's channel at the $q$-th subcarrier; $f_q={qB}/{Q}$ denotes the frequency shift of the $q$-th subcarrier, with $B$ and $Q$ being the bandwidth and subcarrier number, respectively, in the OFDM system; $f_c=c/\lambda$ denotes the carrier
frequency with $c$ being the speed of light. Subsequently, the narrowband channel model in \eqref{r3} can be extended into the wideband scenario, such that at the $q$-th subcarrier of the $k$-th UE, we have
\begin{equation}
\mathbf{h}_{k,q}\in \mathbb{C}^{M} =\sum\limits_{l=1}^{L}{\alpha_{l,k}e^{-j{2\pi} f_q\tau_{l,k}}} \mathbf{a}(\bm{\Phi}_{l,k,q}).
\label{r4}
\end{equation}
Equivalent to \eqref{r4}, the channel between the $k$-th UE and the $m$-th antenna at the BS over all subcarriers is described as per
\begin{equation}
\mathbf{h}_{k,m}\in \mathbb{C}^{Q} =\sum\limits_{l=1}^{L}{\alpha_{l,k}} e^{-j\frac{2\pi}{\lambda}\mathbf{s}_m^T{\cdot\bm{\Theta}}_{l,k}}\Big( \mathbf{b}(\tau_{l,k})\circ \mathbf{d}(\mathbf{s}_m)\Big),
\label{rr4}
\end{equation}
where $\mathbf{b}(\tau_{l,k}) = [e^{-j2\pi f_1{\tau _{l,k}}}, \cdots ,{e^{-j2\pi f_Q{\tau _{l,k}}}}]^T $ is the frequency steering vector in narrowband scenarios that points towards the time delay $\tau_{l,k}$. The term $\mathbf{d}(\mathbf{s}_m)  \!=\! [e^{-j\frac{2\pi}{c}f_1 \mathbf{s}_m^T{\cdot\bm{\Theta}}_{l,k}},$ $ \cdots ,e^{-j\frac{2\pi}{c}f_Q \mathbf{s}_m^T{\cdot\bm{\Theta}}_{l,k}}]^T$ is related to the beam squint effect.
\begin{algorithm}[tbp]
\caption{APU for Geometric Shaping}
\begin{algorithmic}[1]
\label{GS}
\REQUIRE $\mathbf{P}_0\in \mathbb{C}^{T^2M\times K}$, $\bm{{\rm{H}}}_0\in \mathbb{C}^{M\times K}$, and $N$.
\FOR{$n=0,1,\cdots,N$}
   \STATE Update $i_n$ via \eqref{q2one}.
   \STATE Update $j_n$ via \eqref{equal2}.
   \STATE Update $\mathbf{H}_{n+1} = \mathbf{H}_n^{(i_n,j_n)}$ and $\mathbf{P}_{n+1} =\mathbf{P}^{(j_n)}_n$.
   \STATE  {\textbf{Stop}} {if the termination condition $C_2(\mathbf{H}_{n+1}) < C_2(\mathbf{H}_n)$ is satisfied.}
\ENDFOR
\ENSURE $\mathbf{H}_n$.
\end{algorithmic}
\label{alg1}
\end{algorithm}
\begin{algorithm}[t]
\caption{PGA for Amplitude Tapering}
\label{algpga_taper}
\begin{algorithmic}[1]
\REQUIRE $\mathbf{H}\in \mathbb{C}^{M\times K}$, $P$, $N$, $\delta_1$, $\delta_2$, $\beta$, and $c$.
\STATE {\textbf {Initialize}}: $\mathbf{a}_0$ and $\alpha_0>0$. 
\FOR{$n=0,1,\ldots,N$}
    \IF{$\|\nabla_{\mathbf{a}_n} C_3(\mathbf{H},\mathbf{a}_n)\|_2\le \delta_1$}
        \STATE \textbf{break}
    \ENDIF
    \STATE $\alpha = \alpha_0$  (Initialization in backtracking line search).
    \REPEAT
        \STATE $\tilde{\mathbf{a}} = \Pi_{\mathcal{F}}(\mathbf{a}_n+\alpha\,\nabla_{\mathbf{a}_n} C_3(\mathbf{H},\mathbf{a}_n))$. 
        \STATE $\alpha  = \beta\,\alpha$.    \UNTIL{$C_3(\mathbf{H},\tilde{\mathbf{a}})\ge C_3(\mathbf{H},\mathbf{a}_n) + c\,\alpha\,\|\nabla_{\mathbf{a}_n} C_3(\mathbf{H},\mathbf{a}_n)\|_2^2$ \textbf{or} $\alpha<\delta_2$}.\STATE $\mathbf{a}_{n+1} =  \tilde{\mathbf{a}}$.
\ENDFOR
\ENSURE $\mathbf{A}_n = D(\mathbf{a}_n)$.
\end{algorithmic}
\end{algorithm}
\par Note that on the $l$-th path of the $k$-th UE's channel, the beam squint angle at the $q$-th subcarrier and the $m$-th antenna can be measured by
\begin{equation}
\Delta^{(m,q)}_{l,k}\!=\!\Big|\angle [\mathbf{a}(\bm{\Theta}_{l,k})]_m - \angle [\mathbf{a}(\bm{\Phi}_{l,k,q})]_m \Big| 
\!=\!\frac{2\pi f_q}{c}\,|\mathbf{s}_m^T\cdot\bm{\Theta}_{l,k}|,
\label{v1}
\end{equation}
where $\mathbf{a}(\bm{\Theta}_{l,k})$ and $\mathbf{a}(\bm{\Phi}_{l,k,q})$ are defined in \eqref{f2} and \eqref{g3}, respectively. Furthermore, the average beam squint angle across all subcarriers at the $m$-th antenna and across all antennas at the $q$-th subcarrier can be respectively quantified by
\begin{numcases}{}
\Delta^{(m)}_{l,k} = \frac{1}{Q}\sum\nolimits_{q=1}^Q\Delta^{(m,q)}_{l,k}, \ \forall m\in \llbracket M \rrbracket,
\label{vv1}
\\
\Delta^{(q)}_{l,k} = \frac{1}{M}\sum\nolimits_{m=1}^{M}\Delta^{(m,q)}_{l,k}, \ \forall q \in \llbracket Q \rrbracket.
\label{vvq1}
\end{numcases}
According to \eqref{v1}--\eqref{vvq1}, the beam squint angle is determined by two factors: 1) frequency shift of subcarriers $f_q$; 2) position shift of array elements $\mathbf{s}_m$. The former factor indicates that a larger bandwidth $B$, which determines the frequency shift $f_q={qB}/{Q}$ at the $q$-th subcarrier, leads to a more severe squinted beam. Assume without loss of generality that the spatial origin is located at the center of the array. Then, the latter factor indicates that the beam squint becomes more pronounced as an array element moves further away from the center. Unlike the URA, as depicted in Fig. \ref{AllPattern}, typical NUAs with the exception of HuD feature a circular aperture characterized by elements that are more densely arranged, making them more resilient to the beam squint effect.
\begin{figure*}[!t]
\subfigure[Effect of array element number.]
{\includegraphics[height=5.1cm,width=6.01cm]{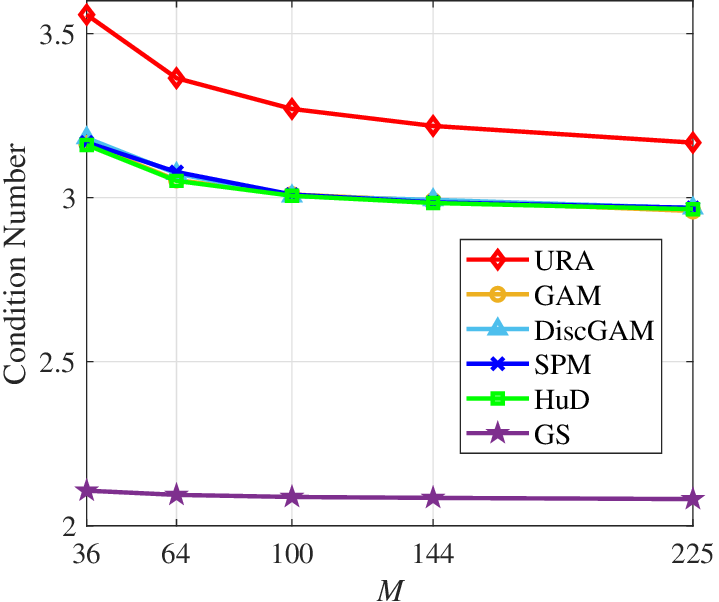}}
\vspace{-0.0cm}
\subfigure[Effect of inter-element spacing.]
{\includegraphics[height=5.1cm,width=6.01cm]{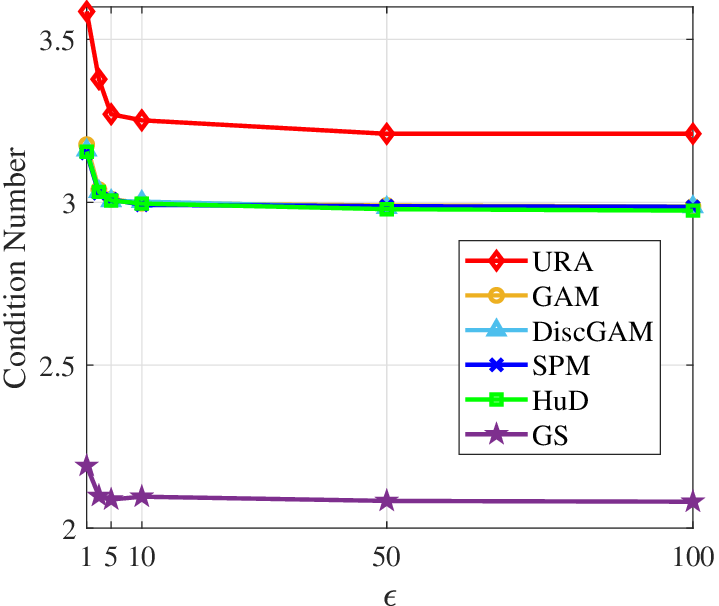}}
\hspace{-0.0cm}
\subfigure[Effect of UE number.]
{\includegraphics[height=5.1cm,width=6.01cm]{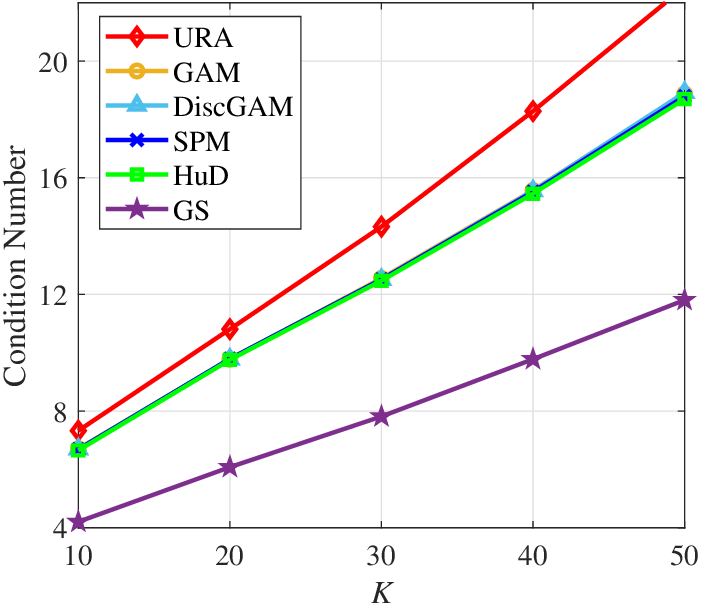}}
\hspace{-0.0cm}
\caption{Comparison of the condition number of the multi-user channel achieved using different array distributions.}
\label{CondNum1}
\end{figure*}
\begin{figure}[!t]
\centerline{\includegraphics[width=7.0cm]{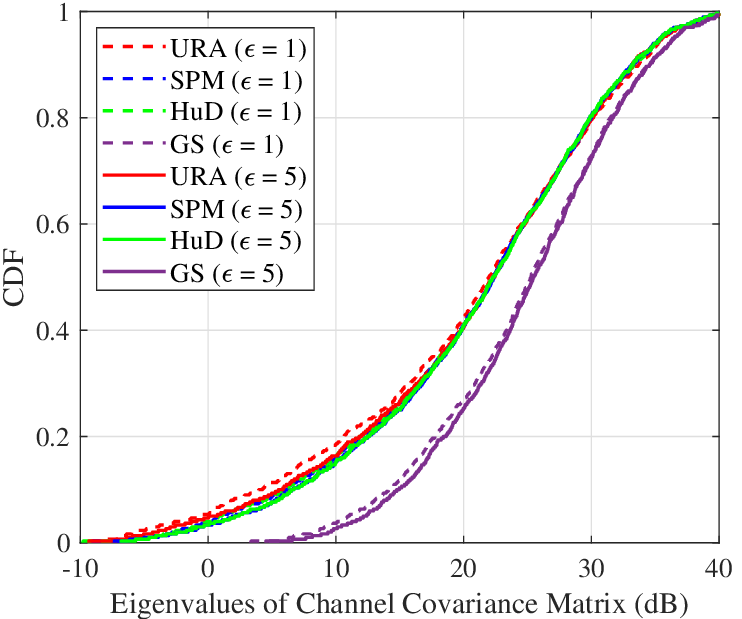}}
\centering
\caption{Comparison of the eigenvalue CDFs for the CCM achieved using different array distributions.}
\label{CondNum2}
\end{figure}
\par Besides the previously mentioned steering vector, the beam squint effect can also be examined via the beam pattern. This pattern offers a visualization of the combined influence of every element within an array. Consider a wideband scenario, where the magnitude of the normalized array factor at the $q$-th subcarrier can be calculated in a similar fashion to \eqref{e80} as\cite{HUDX}
\begin{equation}
\begin{aligned}
|A(\bm{\Phi}_{l,k,q})| = \frac{1}{M} \left|\sum\nolimits_{m=1}^{M} e^{-j\frac{2\pi}{\lambda}\mathbf{s}_{m}^T{\cdot{\bm{\Phi}}}_{l,k,q}}\right|, \ \forall q \in \llbracket Q \rrbracket.
\label{e802}
\end{aligned}
\end{equation}
As per \eqref{e802} and \eqref{g3}, the beam pattern related to the $l$-th path of the $k$-th UE's channel varies with frequency. This implies that different subcarriers in the OFDM system will observe distinct angles of arrival for the same path, while the beam squint degree at the $q$-th subcarrier depends on the frequency shift $f_q$ and the position shift of antennas $\mathbf{s}_m, \forall m\in \llbracket M \rrbracket$.
\section{Experimental Results}
\label{expe}
\begin{figure}[!t]
\centerline{\includegraphics[width=7.0cm]{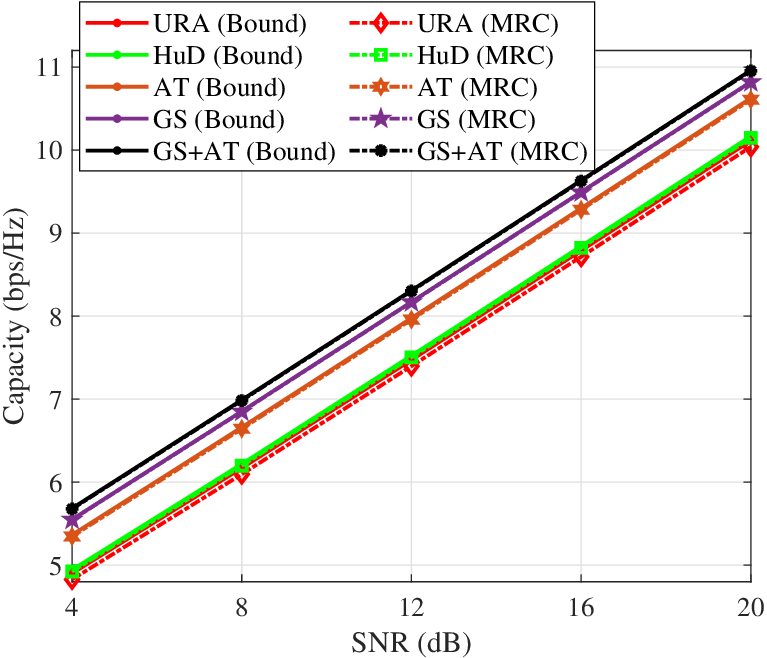}}
\centering
\caption{Comparison of the average per-user capacity of the multi-user channel achieved using different array distributions. }
\label{CapComp1}
\end{figure}

\begin{figure*}[!t]
\subfigure[Effect of element number.]
{\includegraphics[height=5.1cm]{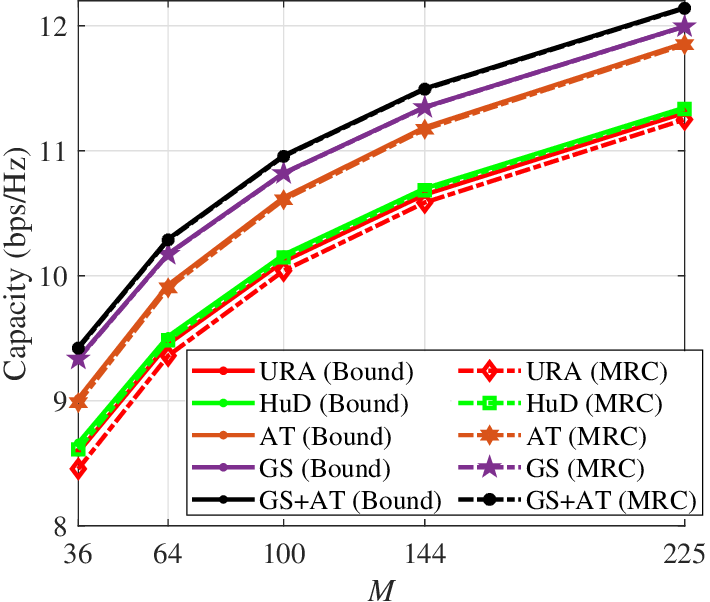}}
\vspace{-0.0cm}
\subfigure[Effect of inter-element spacing.]
{\includegraphics[height=5.1cm]{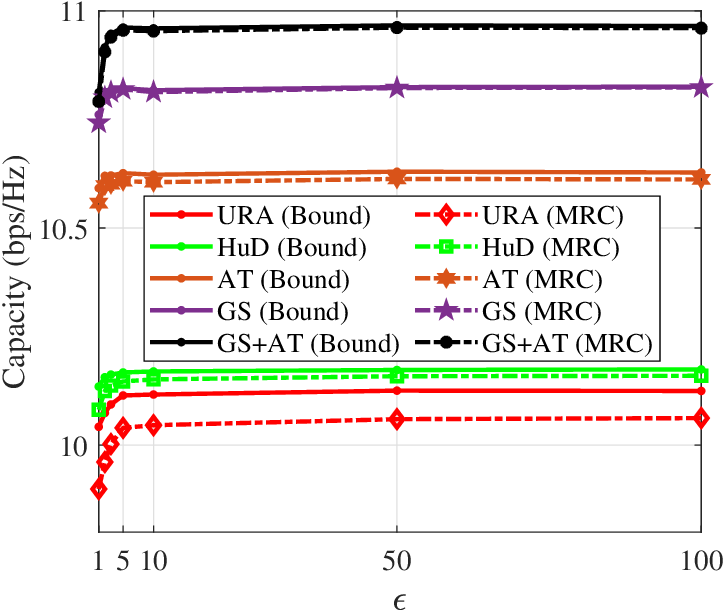}}
\hspace{-0.0cm}
\subfigure[Effect of UE number.]
{\includegraphics[height=5.1cm]{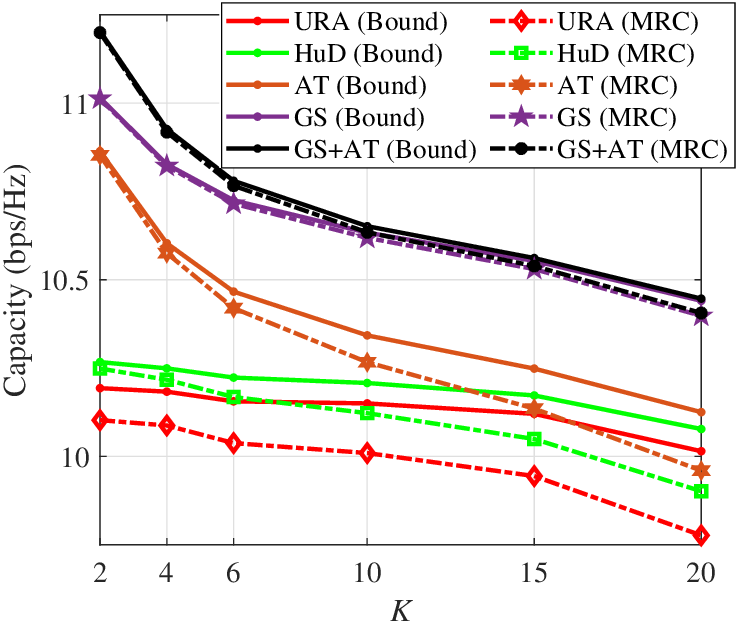}}
\hspace{-0.0cm}
\caption{Comparison of the average per-user capacity of the multi-user channel achieved using different array distributions at an SNR of $20$ dB.}
\label{AVE}
\end{figure*}

\begin{figure}[!tbp]
\centering
{\includegraphics[width=7.0cm]
{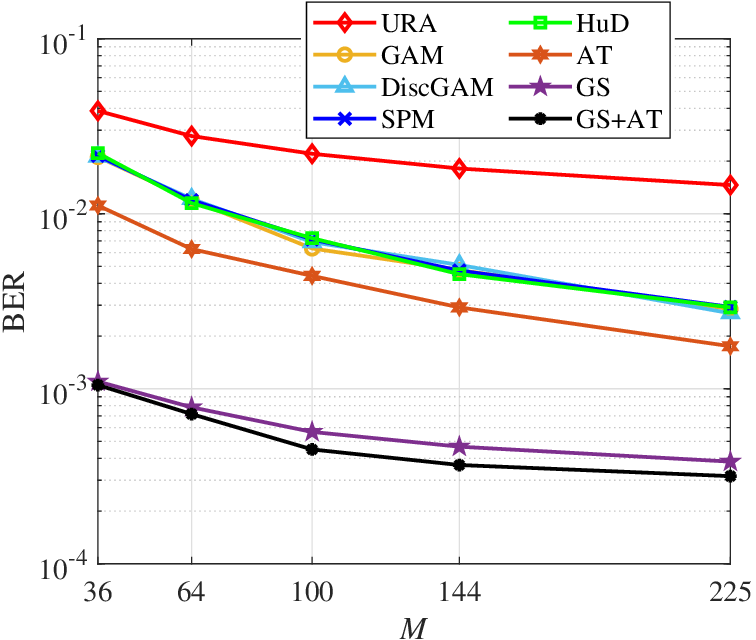}}
\caption{Comparison of the BER achieved using different array distributions at an SNR of $20$ dB.}
\label{BERULDL}
\end{figure}

This section evaluates the performance of several NUAs using GAM, DiscGAM, SPM, HuD, and geometrically shaped distributions, respectively. Their performance is also compared with that of a URA with the same aperture size and element number. The URA or NUA is deployed at the BS to serve multiple single-antenna UEs. In the OFDM system, the wavelength $\lambda$ corresponds to a carrier frequency of $f_c=6$ GHz, and the tested subcarrier has a frequency shift of $f_s=1$ MHz. For comparison, the curves labeled “GS”, “AT”, and “GS+AT” in Figs. \ref{CondNum1}-\ref{BERULDL} correspond to array optimization results using geometric shaping in Algorithm \ref{GS}, amplitude tapering in Algorithm \ref{algpga_taper}, and joint geometric shaping and amplitude tapering, respectively. The parameter settings of Algorithms \ref{GS} and \ref{algpga_taper} include $T=10$, $N=300$, $P=M$, $\beta=0.5$, $c=10^{-4}$, $\delta_1=10^{-8}$, and $\delta_2=10^{-12}$.
\subsection{Evaluation on Channel Characteristics}
\label{experiment1}
We generate the narrowband channels as described in Section \ref{SignalModel}. The default simulation parameter settings include $M=100$, $L=50$, $K=3$, and $d=\epsilon\lambda$ with a constant $\epsilon=5$. The channel vector $\mathbf{h}_k$ in \eqref{r3} is independently generated for each UE $k$. The channel parameters are randomly and uniformly generated in each of the $500$ Monte-Carlo trials. Specifically, the time delays of multipath components are uniformly and randomly distributed within the interval $(0,500\text{ns}]$. The channel is assumed to consist of one line-of-sight (LOS) path and $(L-1)$ non-line-of-sight (NLOS) paths. The complex gains of the NLOS and LOS paths follow zero-mean complex Gaussian distributions with variances $10^{-4}$ and $0.195$, respectively. Correspondingly, the power ratio between the LOS and NLOS paths, referred to as the $K$-factor, equals $16$ dB\cite{pasic2025millimeter,matolak2016air}. In each trial, we randomly choose one UE and set the azimuth and elevation angles of its LOS path to zero, thereby aligning this path with the main lobe of the array. Meanwhile, the elevation and azimuth angles of all other paths are randomly and uniformly drawn from the intervals $[0,\pi/2]$ and $[0,\pi]$, respectively.
\par Based on the generated channel matrix $\mathbf{H}\in\mathbb{C}^{M\times K}$, we evaluate its orthogonality in terms of the condition number. In general, a more orthogonal channel corresponds to a condition number approaching 1. Figs. \ref{CondNum1}(a)-(c) present the condition number results versus array element number $M$, inter-element spacing $d=\lambda\epsilon$, and UE number $K$, respectively. Specifically, Fig. \ref{CondNum1}(a) varies $M$, while fixing $\epsilon$ and $K$, to test its impact on the condition number. Fig. \ref{CondNum1}(a) shows that channel orthogonality gradually improves as the number of array elements increases, confirming the advantages of massive MIMO systems. Compared with the URA, all NUAs achieve lower condition numbers for different numbers of array elements. This phenomenon can be attributed to the presence of numerous grating lobes in the URA, which cause interference among the channel vectors of different UEs and consequently reduce the orthogonality of the channel matrix. Furthermore, the NUA with a geometrically shaped distribution using Algorithm \ref{GS} achieves the best condition number results, reflecting the positive correlation between channel capacity and orthogonality. Unlike the GAM, DiscGAM, SPM, and HuD distributions, which are fixed for a given number of array elements, the geometrically shaped distribution via Algorithm \ref{GS} adapts to the channel state information, enhancing channel capacity and yielding more favorable propagation characteristics. A similar trend in the condition number comparison is observed in Fig. \ref{CondNum1}(b), where the proposed GS method achieves the best performance. Fig. \ref{CondNum1}(c) investigates the impact of the number of UEs on channel orthogonality. The results indicate that as the number of UEs served by the BS increases, the probability of beam overlapping and the resulting inter-user interference also increase. Moreover, the URA suffers from spatial aliasing artifacts, which further degrade channel orthogonality when the number of UEs becomes large. In the challenging scenario where $K=50$, the NUA with a geometrically shaped distribution significantly outperforms the URA in Fig. \ref{CondNum1}(c), demonstrating enhanced channel orthogonality.
\par In addition to the condition number of the channel matrix, we evaluate the channel characteristics using another performance metric, namely the cumulative distribution functions (CDF) of the eigenvalues of CCM. An orthogonal channel typically exhibits stable behavior, with the eigenvalue distribution of its CCM displaying low variance and narrowly spread CDFs. Fig. \ref{CondNum2} shows that NUAs achieve superior CDF performance compared to URAs for inter-element spacings of $d=\epsilon\lambda$, with $\epsilon=1$ and $\epsilon=5$ tested. The joint analysis in Figs. \ref{CondNum1} and \ref{CondNum2} clearly highlights the advantages of NUAs over URAs in terms of favorable propagation.
\subsection{Evaluation on System-Level Performance}
\label{systemeva}
We conduct the system-level evaluation of array distribution designs using the channel capacity criterion. The experimental setup follows that in Section \ref{experiment1}. Fig. \ref{CapComp1} presents the average per-user capacity of the multi-user channel $\mathbf{H}\in\mathbb{C}^{M\times K}$ versus signal-to-noise ratio (SNR). Note that the capacity bound in Fig. \ref{CapComp1} refers to the value $C_3(\mathbf{H},\mathbf{a})$ in \eqref{cap3} for the AT and GS+AT methods, while referring to the value $C_2(\mathbf{H})$ in \eqref{cpc} for all other methods. The MRC capacity in Fig. \ref{CapComp1} refers to the value $C_1(\mathbf{H})$ in \eqref{cpc0} under the ZF beamforming scheme for all methods except the AT and GS+AT methods. Specifically, the MRC capacity for the AT and GS+AT methods is defined similarly to \eqref{cpc0} and \eqref{SINR} by replacing the $k$-th UE's channel $\mathbf{h}_k$ with its amplitude tapering version $D(\mathbf{a})\cdot \mathbf{h}_k$ for all $k\in \llbracket K \rrbracket$.
\par Fig. \ref{CapComp1} shows that NUAs consistently outperform the URA under different SNR conditions. Additionally, the GS method performs better than the AT method, which indicates the more significant impact of optimizing the antenna array distribution compared with the impact of the excitation amplitudes on channel capacity. Moreover, Fig. \ref{CapComp1} demonstrates that the joint use of amplitude tapering and geometric shaping can ensure improvement in both aspects. As shown, the proposed GS+AT method achieves the best performance, yielding a capacity gain of approximately 0.8 bps/Hz compared with the URA. 
\par Next, we examine the effects of different scenario parameters, including $M$, $\epsilon$, and $K$, on the channel capacity. In Fig. \ref{AVE}(a), with  $\epsilon$ and $K$ fixed, we vary $M$ to evaluate the average per-user capacity of the multi-user channel $\mathbf{H}\in\mathbb{C}^{M\times K}$. Similarly, Figs. \ref{AVE}(b) and (c) vary $\epsilon$ and $K$, respectively, while keeping the other two parameters constant, to report the capacity results. As shown in Figs. \ref{AVE}(a)-(c), the proposed GS+AT method consistently achieves the best performance across different scenario parameter settings. Furthermore, the MRC capacity attained by the GS+AT method closely approaches its capacity bound. This reflects a positive correlation between maximizing the capacity bound via the GS+AT method and enhancing the MRC capacity.
\par In addition to the capacity metric, we also compare different array distributions using the BER metric. Fig. \ref{BERULDL} presents BER results for uplink communication employing the quaternary phase shift keying (QPSK) modulation and a matched filter detector. As mentioned in Section \ref{FavPro}, favorable propagation emerges when the multi-user channel approaches an orthogonal matrix and exhibits stable behavior in the singular value distribution. This property enables the matched filter detector asymptotically optimal as the number of antennas increases \cite{xiao2009improved,yang2013performance}, making it increasingly attractive for massive
MIMO systems. As shown in Fig. \ref{BERULDL}, NUAs achieve substantial BER improvements over the URA, attributed to their enhanced favorable propagation. Notably, the proposed GS+AT method yields the best BER performance across scenarios with different numbers of antennas.
\subsection{Evaluation on Beam Squint}
We compare the performance of URA and NUAs regarding the beam squint effect in a wideband scenario. The experimental setup described in Section \ref{experiment1} is adopted, except that the inter-element spacing is set to half-wavelength in order to focus on the assessment of beam squint without the influence of grating lobes in the URA. We consider a wideband scenario with a bandwidth $B = 1$ GHz and $Q=1024$ subcarriers. The parameters of elevation angle $\theta_{l,k}$, azimuth angle $\varphi_{l,k}$, and time delay $\tau_{l,k}$ for the $l$-th path of the $k$-th UE's channel are configured as $30^\circ$, $30^\circ$, and 100 ns, respectively. 
\par
\begin{figure}[!t]
\centerline{\includegraphics[width=7.1cm]{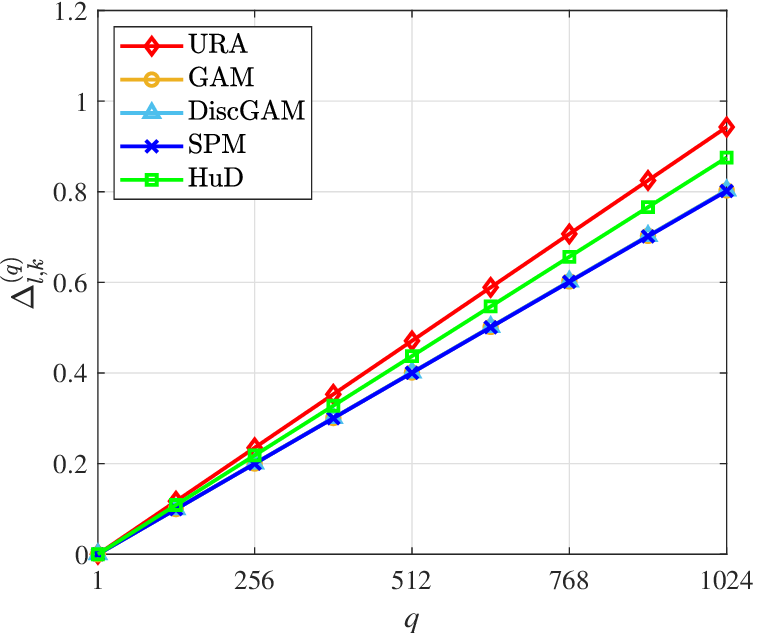}}
\centering
\caption{Comparison of the average angle of beam squint across all antennas at the $q$-th subcarrier, represented by $\Delta^{(q)}_{l,k}$, for all $q\in \llbracket Q \rrbracket$ with $Q=1024$.}
\label{BS2}
\end{figure}

\begin{figure}[!t]
\centerline{\includegraphics[width=7.1cm]{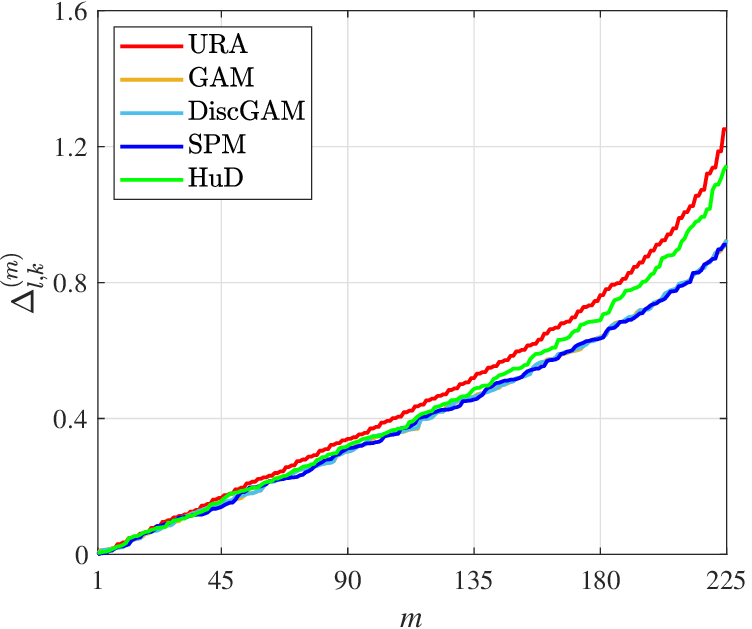}}
\centering
\caption{Comparison of the average angle of beam squint across all subcarriers at the $m$-th antenna, represented by $\Delta^{(m)}_{l,k}$, for all $m\in \llbracket M \rrbracket$ with $M=225$.}
\label{BS3}
\end{figure}

 As mentioned in Section \ref{beamsquint}, the beam squint angle is influenced by two factors, namely, the frequency shift of subcarriers from the carrier frequency, and the position shift of antennas from the center of the array. To quantify the relationship between beam squint and these two factors, we use \eqref{vv1} and \eqref{vvq1} to measure the average angle of beam squint across all subcarriers and all antennas, respectively. Figs. \ref{BS2} and \ref{BS3} verify that a larger shift in frequency or position corresponds to a more severe squinted beam. Furthermore, HuD outperforms URA in terms of the average angle of beam squint in Figs. \ref{BS2} and \ref{BS3}. Nevertheless, HuD falls behind other typical NUAs, including GAM, DiscGAM, and SPM, which feature circular apertures characterized by elements arranged in a denser manner. The results in Figs. \ref{BS2} and \ref{BS3} demonstrate that a more compact distribution of elements within the array achieves enhanced robustness to beam squint for a given bandwidth.
\section{Conclusion and Future Work}
\label{concl}
In this paper, we proposed  NUAs with large inter-element spacing for massive MIMO systems. Several constellation-inspired NUA designs, including HuD, GAM, DiscGAM, and SPM distributions, were introduced by drawing inspiration from mutual-information-optimal constellation shaping in the signal domain. Experimental results  demonstrated the superiority of NUAs over the traditional URA in terms of grating lobe suppression, aperture efficiency, favorable propagation, and channel capacity. In the wideband scenario, NUAs exhibit greater robustness against beam squint due to their more compact spatial distribution. A key contribution of this work is the EMIT framework, which establishes a principled connection between the spatial sampling pattern of an antenna array and the capacity of MIMO channels. By recognizing that the beam pattern is an invertible Fourier transform of the spatial element distribution, we formulated the array design as a capacity maximization problem over both the array geometry and the excitation profile. This gave rise to two complementary shaping strategies, amplitude tapering and geometric shaping, and their joint optimization. Numerical results confirmed that the proposed EMIT-based shaping schemes yield substantial capacity gains beyond those achieved by the NUA geometry alone.
\par Several promising directions remain for future work. First, the EMIT framework proposed in this work represents an initial step toward information-theoretic array optimization; extending it to more general channel models  and NUA patterns warrants further investigation. Second, NUAs hold significant potential for emerging applications such as integrated sensing and communications (ISAC), cell-free massive MIMO, and non-terrestrial networks (NTN), where irregularly spaced antenna deployments are inherent to the system architecture. Third, the analogy between spatial sampling and constellation shaping opens the possibility of jointly optimizing antenna geometry and modulation under a unified information-theoretic objective. Finally, experimental validation on hardware testbeds would provide valuable insights into practical applications of NUAs.
\bibliographystyle{IEEEtran}
\bibliography{Bibliography}
\end{document}